\documentclass[fleqn,usenatbib]{mnras}

\usepackage{newtxtext,newtxmath}
\usepackage[T1]{fontenc}
\usepackage{lipsum}
\usepackage{lineno}
\usepackage{bm}

\DeclareRobustCommand{\VAN}[3]{#2}
\let\VANthebibliography\thebibliography
\def\thebibliography{\DeclareRobustCommand{\VAN}[3]{##3}\VANthebibliography}

\usepackage{graphicx}	% Including figure files
\usepackage{amsmath}	% Advanced maths commands
\usepackage[dvipsnames]{xcolor}
\newcommand{\high}{\textsf{High}}
\newcommand{\highbase}{\textsf{Highbase}}
\newcommand{\cph}{\textsf{c000\_ph100}}
\newcommand{\rockstar}{{\sc Rockstar}}
\newcommand{\compaso}{{\sc CompaSO}}
\newcommand{\abacus}{{\sc Abacus}}
\newcommand{\summit}{{\sc AbacusSummit}}
\newcommand{\cadence}{{\sc HighCadence}}

\title[AbacusSummit merger trees]{Constructing high-fidelity halo merger trees in \summit{}}

\author[S. Bose et al.]{
Sownak Bose,$^{1,2}$\thanks{E-mail: sownak.bose@durham.ac.uk (SB)}
Daniel J. Eisenstein,$^{1}$
Boryana Hadzhiyska,$^{1}$
Lehman H. Garrison,$^{3}$
\newauthor{\,and Sihan Yuan$^{1,4}$}
\\
$^{1}$Center for Astrophysics | Harvard \& Smithsonian, 60 Garden St., Cambridge, MA 02138, USA\\
$^{2}$Institute for Computational Cosmology, Department of Physics, Durham University, Durham DH1 3LE, UK\\
$^{3}$Center for Computational Astrophysics, Flatiron Institute, 162 Fifth Avenue, New York, NY 10010, USA\\
$^{4}$Kavli Institute for Particle Astrophysics and Cosmology, Stanford University, Stanford, CA 94305, USA
}

\date{Accepted XXX. Received YYY; in original form ZZZ}

\pubyear{2022}

\begin{document}
\label{firstpage}
\pagerange{\pageref{firstpage}--\pageref{lastpage}}
\maketitle

% Abstract of the paper
\begin{abstract}
Tracking the formation and evolution of dark matter haloes is a critical aspect of any analysis of cosmological $N$-body simulations. In particular, the mass assembly of a halo and its progenitors, encapsulated in the form of its merger tree, serves as a fundamental input for constructing semi-analytic models of galaxy formation and, more generally, for building mock catalogues that emulate galaxy surveys. We present an algorithm for constructing halo merger trees from \summit{}, the largest suite of cosmological $N$-body simulations performed to date consisting of nearly 60 trillion particles, and which has been designed to meet the Cosmological Simulation Requirements of the Dark Energy Spectroscopic Instrument (DESI) survey. Our method tracks the cores of haloes to determine associations between objects across multiple timeslices, yielding lists of halo progenitors and descendants for the several tens of billions of haloes identified across the entire suite. We present an application of these merger trees as a means to enhance the fidelity of \summit{} halo catalogues by flagging and ``merging'' haloes deemed to exhibit non-monotonic past merger histories. We show that this cleaning technique identifies portions of the halo population that have been deblended due to choices made by the halo finder, but which could have feasibly been part of larger aggregate systems. We demonstrate that by cleaning halo catalogues in this post-processing step, we remove potentially unphysical features in the default halo catalogues, leaving behind a more robust halo population that can be used to create highly-accurate mock galaxy realisations from \summit{}.  
\end{abstract}

% Select between one and six entries from the list of approved keywords.
% Don't make up new ones.

%%%%%%%%%%%%%%%%%%%%%%%%%%%%%%%%%%%%%%%%%%%%%%%%%%

%%%%%%%%%%%%%%%%% BODY OF PAPER %%%%%%%%%%%%%%%%%%
%\linenumbers

\section{Introduction}
\label{sec:intro}

In our present paradigm of structure formation, galaxies are thought to form within potential wells of dark matter that have collapsed out of density fluctuations in the primordial Universe due to gravitational instability. These potential wells -- so-called `haloes' of dark matter -- are self-gravitating, virialised structures that have decoupled from the background expansion of the Universe. As gas shock heats and, subsequently, cools and condenses within dark matter haloes, it sparks star formation that leads to the eventual formation of galaxies.

Modelling the hierarchical formation and evolution of dark matter haloes is therefore a fundamental component of any theoretical framework of galaxy formation. A well-known, early approach to this problem was presented in the analytic model of \cite{Press1974} who, assuming initial fluctuations seeded by a Gaussian random field, derived a methodology for computing the multiplicity function of dark matter haloes as a function of mass and redshift. Despite the relative simplicity of the model, the number density of haloes predicted by the Press-Schechter approach has been shown to be in reasonable agreement with early numerical experiments of cosmic structure formation \citep[e.g.][]{Efstathiou1988,Lacey1994}. 

A refinement of this model, often dubbed the `extended' Press-Schechter (EPS) theory, was later introduced by \cite{Bond1991}, who described a framework for computing the distribution of haloes with mass $M$ at some redshift $z$ by following trajectories of individual mass elements through the linear overdensity field. In particular, EPS enables the calculation of the quantity $f\left(M_1,z_1 \vert M_2,z_2\right)$, or the fraction of mass from haloes with mass $M_2$ at redshift $z_2$ that are contained in {\it progenitor} haloes of mass $M_1$ identified at an earlier redshift $z_1$. The EPS method therefore enables the computation of halo formation and merger rates over multiple epochs \citep[e.g.][]{Lacey1993,Somerville1999,Parkinson2008} -- in other words, the formation and evolution of progenitor and descendant haloes, the ensemble of which defines a {\it halo merger tree}. Once constructed, the merger tree acts as the backbone upon which semi-analytic models of galaxy formation may be built \citep[e.g.][]{Kauffmann1993,Cole2000,Croton2006,Somerville2008,Benson2012,Henriques2015,Lagos2018}.

While analytic methods like EPS are instructive, modelling the formation and evolution of cosmic structures self-consistently in their large-scale environment is the order of the day if we wish to compare theoretical models to the observed Universe. Over the last several decades, cosmological $N$-body simulations have become near indispensable tools for augmenting our understanding of structure formation. In terms of the computational volume of their calculations, simulations that model the dark matter only are currently the industry standard \citep[e.g.][]{Springel2005,Klypin2011,Heitmann2015,Potter2017,Garrison2018, Elahi2018,Baugh2019,Ishiyama2020}. Halo catalogues and merger trees extracted from these simulations may then be augmented with models of the galaxy-halo connection to build mock catalogues that can then be compared to galaxy surveys \citep[see][for a review]{Wechsler2018}. Although hydrodynamical simulations of galaxy formation are continuously improving in both their scale and their sophistication \citep[see][for a review]{Vogelsberger2020}, at present they fall short of the demands of modern galaxy surveys in terms of box size and the number of simulations available. 

 The identification of dark matter haloes in $N$-body simulations is by no means a `solved' problem. At some level, any halo finder has to ultimately resort to making a series of choices when defining the existence and extent of haloes. For example, the first choice comes in deciding whether halo particle memberships are to be determined in configuration space, as in the `friends-of-friends' \citep[][]{Davis1985} or spherical overdensity-based methods \citep[e.g.][]{Press1974,Klypin1997}, in phase space \citep[e.g.][]{Behroozi2013}, or overall characterisations of the cosmic density field \citep[e.g.][]{Neyrinck2005,Falck2012}. A subsequent step may (or may not) involve determining which subsets of `halo' particles are in fact gravitationally bound \citep[e.g.][]{Springel2001,Knollmann2009}. Each choice can lead to slightly different populations of haloes realised from the same particle field \citep[see][for an exhaustive study of the effects of halo finders on the properties of the final catalogue]{Knebe2011}.

Given these uncertainties, halo merger trees may then also be used as useful diagnostic tools for assessing the fidelity of the halo population itself. The dynamical history of these objects introduces further complications as they may `split', fly-by one another, merge partially, or simply flicker above and below the mass threshold for recording a halo. Each of these processes can significantly alter the mass associated with the halo over time. This is particularly significant for applications that use the properties of haloes to build mock galaxy catalogues. A prominent example is that of the Halo Occupation Distribution \citep[HOD, e.g.][]{Peacock2000,Benson2000,Berlind2002,Zheng2005}, in which the probability that a halo contains a certain population of galaxies is dependent on properties like its mass or maximum circular velocity. The dynamical histories of haloes may be stochastic enough to make their final mass very different from the mass they would have had when a galaxy in the real Universe would have formed in them, in which case a model like the HOD may result in a biased mock galaxy population. In this regard, a working definition of a ``well-defined halo'' may simply be a persistent entity -- i.e., that which can be {\it cleanly} associated from one timestep of the simulation to the next. By traversing the merger trees of individual haloes, one can then flag and remove objects that fail to pass this persistence test in post-processing. 

There are now several approaches in the literature for constructing merger trees from cosmological simulations, differing in their complexity and the information content that acts as input for building these trees. While some algorithms attempt to establish associations between haloes across timesteps based simply on their mass/position/velocity in the box \citep[][]{Onions2012}, other methods aim at connecting haloes (or subhaloes) that maximise a merit function based on overlapping particle sets across two timeslices \citep[][]{Klimentowski2010,Elahi2018}. Particle-based correlator methods are the more established flavour in recent times, seeing as they display greater stability in the properties of their merger trees \citep[see, e.g.,][]{Srisawat2013}. Some algorithms assign greater weight in the merit function to those particles that are more tightly bound \citep[based on e.g. their binding energy as in][]{Springel2005,Jiang2014}. Finally, there are also those algorithms that, in addition to using particles to associate haloes, also make use of the past history of individual objects (such as their bulk motion, or indeed their particle membership) to construct the final merger tree \citep[e.g.][]{Behroozi2013b,Han2018}.

As modern galaxy surveys like the Dark Energy Spectroscopic Instrument \citep[DESI,][]{Levi2013}, {\it Euclid} \citep{Laureijs2011}, {\it The Nancy Grace Roman Space Telescope} \citep{Spergel2015}, and the Vera Rubin Observatory's Legacy Survey of Space and Time \citep[LSST,][]{Ivezic2019} continue to ramp up the statistical power of observational data, there is an increasingly pressing demand on theoretical models to keep up to pace. The computational challenge lies in the fact that the relevant numerical simulations not only require the several Gpc$^3$ in volume to emulate these surveys, but they need to simultaneously resolve the mass scales of interest. For example, the largest galaxy sample in DESI will target emission line galaxies (ELGs) in the redshift range between $z\sim0.6-1.6$. Resolving the host haloes of ELGs (typically of the order of $10^{11} {\rm M}_\odot/h$ in halo mass, \citealt{GonzalezPerez2018,Hadzhiyska2020}) with at least a modest number of particles within Gpc-sized boxes requires tens of billions of resolution elements. In this paper, we will describe our methodology for constructing and applying halo merger trees from the \summit{} simulations \citep{Maksimova2021}, which have been designed to meet exactly these specifications.

The layout of this paper is as follows. In Section~\ref{sec:methodology}, we describe the details of the \summit{} suite, and the algorithms used to identify haloes and construct merger trees from them. In Section~\ref{sec:cadence}, we present tests of the robustness of some typical merger tree statistics against the cadence of simulation outputs used to construct them. Section~\ref{sec:applications} presents our main results, and displays some of the primary applications of the merger trees, including the methodology we use to `clean' halo catalogues by coalescing haloes that have been over-split by the halo finder, or undergone any number of the dynamical processes described above. We then demonstrate the power of this cleaning technique by applying `cleaned' \summit{} haloes to a realistic application of constructing mock galaxy catalogues to observed galaxy clustering data. Finally, a summary of our main findings is presented in Section~\ref{sec:conclusions}.

\section{Methodology}
\label{sec:methodology}

In this section, we describe \summit{}, an extremely large set of numerical simulations that are designed to match DESI's cosmological simulation requirements (Section~\ref{sec:summit}), and \compaso{} \citep{Hadzhiyska2021}, the on-the-fly halo finding algorithm that has been specially designed for this project (Section~\ref{sec:compaso}). We then discuss the design philosophy underpinning our merger tree algorithm (Section~\ref{sec:philosophy}), before discussing the methodology itself in detail (Section~\ref{sec:algorithm}). Finally, we conclude by discussing some of the optimisations that have been implemented to speed-up the construction of merger trees in \summit{} (Section~\ref{sec:optimisation}).

\subsection{The \summit{} simulation suite}
\label{sec:summit}

\summit{} is a suite of very large, high-accuracy cosmological simulations of structure formation that have been designed to match -- and exceed -- the Cosmological Simulation Requirements of the DESI survey. In total, we have performed simulations totalling almost 60 trillion particles spanning 97 cosmological models, centred around the \cite{Planck2018} parameters, which acts as our fiducial cosmological model. A subset of these runs also model the evolution of structure in the presence of massive neutrinos. Individual simulations within the \summit{} suite are labelled according to their cosmology, phase (which are set while generating the corresponding initial conditions), box size, and number of resolution elements. For further details, we refer the reader to the main \summit{} release paper by \cite{Maksimova2021} and the accompanying online documentation\footnote{\url{https://abacussummit.readthedocs.io/en/latest/index.html}}. The different numerical setups adopted in each one of the \summit{} simulations are broadly classified as: 
\begin{itemize}
    \item \textsf{Base}: a 2 Gpc$/h$ box with $6912^3$ particles
    \item \highbase{}: a 1 Gpc$/h$ box with $3456^3$ particles
    \item \high{}: a 1 Gpc$/h$ box with $6300^3$ particles
    \item \textsf{Huge}: a 7.5 Gpc$/h$ box with $8640^3$ particles
    \item \textsf{Hugebase}: a 2 Gpc$/h$ box with $2304^3$ particles
    \item \textsf{Fixedbase}: 1.185 Gpc$/h$ box with $4096^3$ particles with fixed amplitude initial conditions
\end{itemize}
Note that \textsf{Base}, \highbase{} and \textsf{Fixedbase} all have the same mass resolution (around $2\times10^9 {\rm M}_\odot/h$ per particle). On the other hand, \high{} has $6\times$ better mass resolution, while both \textsf{Huge} and \textsf{Hugebase} have $27\times$ worse mass resolution. For the majority of this paper, we will only consider the \highbase{} and \high{} simulations. In both instances, we will present results from simulations run at the base cosmology (\textsf{c000} in our notation) and the same initial phase (\textsf{ph100} in our notation). Collectively, we refer to these simulations as \cph{} throughout the remainder of this paper. The only exception to this is our investigation in Section~\ref{sec:HOD}, where we make use of one \textsf{Base} simulation at \textsf{c000} cosmology, but with phase \textsf{ph000}. A comprehensive tabulation of all simulation variants is listed in the online documentation for \summit{}\footnote{\url{https://abacussummit.readthedocs.io/en/latest/cosmologies.html}}. 

All simulations were performed at the Summit supercomputer at the Oak Ridge Leadership Computing Facility, using the \abacus{} code \citep[][see also \citealt{Garrison2019}]{Garrison2021b}. \abacus{} is a high-performance, high-accuracy cosmological $N$-body code that has been optimised for GPU architectures and for large-volume, moderately clustered simulations. By harnessing the power of GPUs, \abacus{} is able to clock over 30 million particle updates per second on commodity dual-Xeon, dual-GPU computers and nearly 70 million particle updates per second on each node of the Summit supercomputer. Importantly, the speedup achieved does not compromise on accuracy: \abacus{} reports a median force accuracy below $10^{-5}$. In-depth descriptions of the near- and far-field force computations, optimisation routines, and code tests are presented in \cite{Garrison2021b}.

\subsection{Halo finding with \compaso{}}
\label{sec:compaso}

To identify haloes in the \summit{} simulations, we have developed a new, on-the-fly halo finder, \compaso{}, which is a hybrid friends-of-friends/spherical overdensity-based halo finder. The algorithm is described in detail in \cite{Hadzhiyska2021}; here, we provide a short summary of its basic operation:
\begin{enumerate}
    \item For each particle in the simulation, we compute its ``local density'', $\Delta$, using a weighting kernel defined as:$$W(r; b)=1-r^2/b^2,$$ where $b$ is set to be equal to 0.4 times the mean interparticle spacing, $l_{{\rm mean}}$. $\Delta$ here is defined in units of the cosmic mean density. 
    \item We then mark particles as being eligible to be in groups if they satisfy the condition $\Delta>60$. These particles are then segmented into what we call `L0 haloes' using a standard friends-of-friends algorithm with linking length $l_{{\rm FoF}}=0.25l_{{\rm mean}}$.
    \item Next, within each L0 halo, we define `L1 haloes' using {\it Competitive Assignment to Spherical Overdensities} (\compaso{}). The particle with the largest $\Delta$ is selected to be the first halo nucleus. We then search outward to find the innermost radius within which the enclosed density drops below the L1 threshold density, $\Delta_{{\rm L1}}$. Particles interior to this radius, $R_{{\rm L1}}$, are tentatively assigned to the L1 group defined by the first nucleus. The set of particles that are interior to 80\% of $R_{{\rm L1}}$ are marked as ``ineligible'' to be future halo nuclei; the remaining particles may yet be eligible.
    \item The remaining ``eligible'' particles are then searched to find the particle with the next highest $\Delta$ that is also a density maximum. This condition is met when a particle is denser that {\it all} other particles -- eligible or not -- that are within a radius of $0.4l_{{\rm mean}}$. If the condition is met, we spawn a new nucleus and once again search outwards for its L1 radius, $R_{{\rm L1}}$, using {\it all} L0 particles.
    \item A particle is assigned to the newly created group if it was previously unassigned {\bf or} if its enclosed density with respect to the new group is at least {\it twice} that of its enclosed density with respect to the group it is currently assigned to. In practice, these enclosed densities are estimated by scaling from $R_{{\rm L1}}$ assuming an inverse square density profile.
    \item We continue searching for new nucleation centres until we reach the minimum density threshold, defined as the central density of a singular isothermal sphere with 35 particles within a radius enclosing 200 times the average background density of the universe.
    \item Finally, within each L1 halo, we repeat steps (iii)-(v) to identify `L2 haloes'; the centre-of-mass of the largest L2 halo is used to define the centre relative to which all L1 halo statistics are outputted. 
\end{enumerate} 
As part of the \summit{} data products, we output properties of all L1 haloes containing at least 35 particles. Conversely, we only store masses for the 5 largest L2 haloes within each L1 halo. Haloes are arranged across 34 separate files (or ``superslabs'') which themselves are comprised of sets of 50 individual slabs which the computational domain in \abacus{} is split into \citep[see][for further details]{Garrison2021b}. \cite{Hadzhiyska2021} present a number of comparisons of the halo populations identified by \compaso{} with those of haloes identified using the \rockstar{} halo finder \citep{Behroozi2013} run on the same particle sets. 

Note that other configuration space-based halo finders perform an additional ``unbinding'' step that is used to discard high-velocity particles from the set assigned to a given halo. In the case of \compaso{}, we have opted to forego such a step: both due to constraints this places on the runtime efficiency of the \abacus{} code, and also because energy-based unbinding algorithms have problems inherently associated with the method. Instead, we apply a ``cleaning'' procedure in post-processing (described in Section~\ref{sec:cleaning}) that is very effective at identifying and filtering haloes with a large fraction of unbound particles. Section 3.3 in \cite{Hadzhiyska2021} presents a series of detailed tests that compares the outcome of our cleaning method with a halo catalogue produced using \compaso{} and unbinding.

\begin{figure*}
    \centering
    \includegraphics[width=\textwidth]{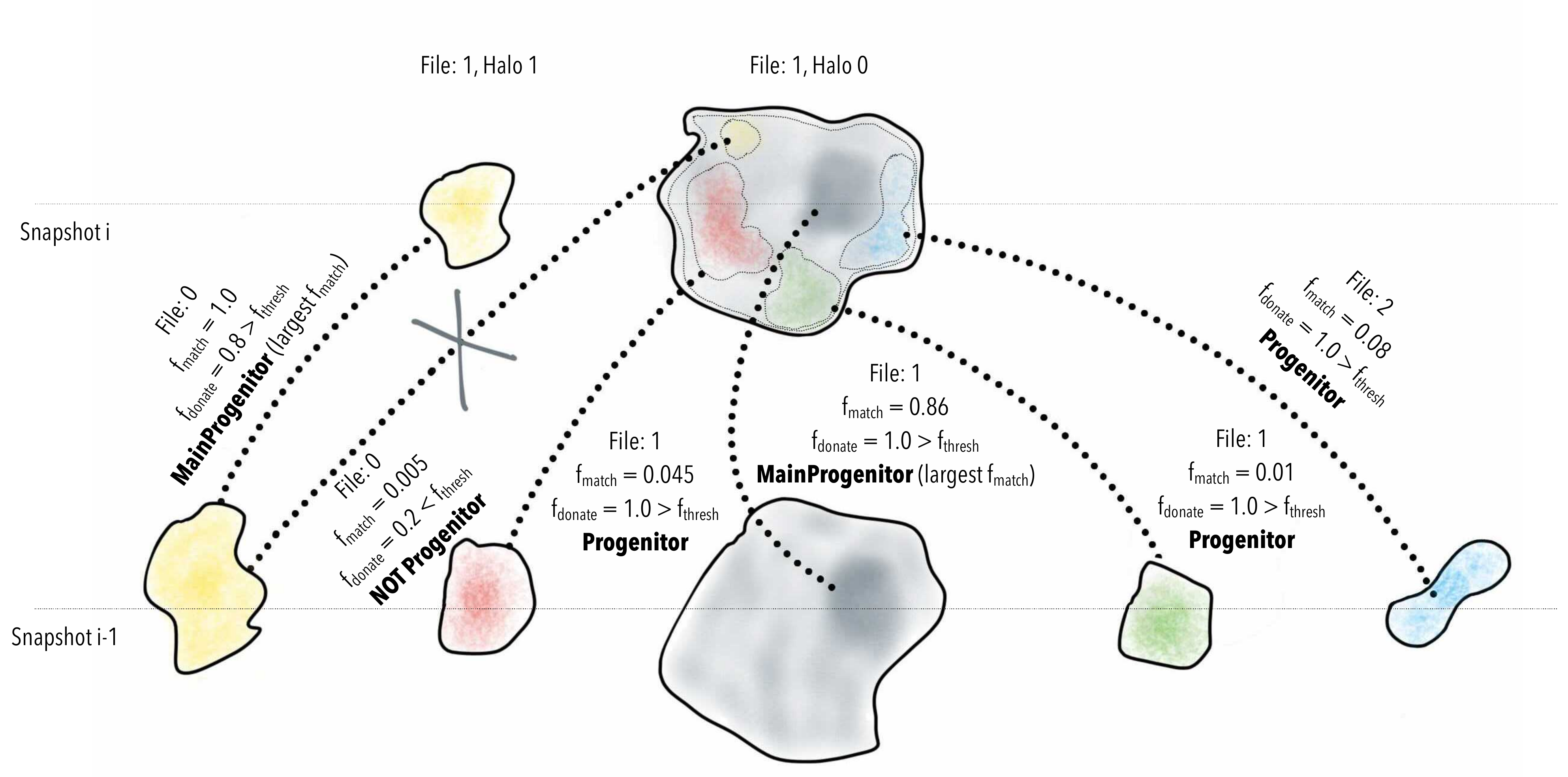}
    \caption{A schematic illustration of the merger tree algorithm. We identify associations between haloes across multiple time slices by tracking subsamples of halo particles that have been tagged. A candidate association is flagged when a halo in snapshot {\tt i-1} donates some fraction of its subsampled particles, $f_{{\rm donate}}$, to a halo in snapshot {\tt i}; these are marked by the dotted lines. The candidate is then marked as a {\tt Progenitor} if $f_{{\rm donate}}\geq f_{{\rm thresh}}$; throughout the \summit{} suite, we assume $f_{{\rm thresh}}=0.5$. The quantity $f_{{\rm match}}$ is defined as the fraction of subsampled particles in a halo in snapshot {\tt i} that was found in a candidate association in snapshot {\tt i-1}. The association with the largest $f_{{\rm match}}$ is marked as the {\tt MainProgenitor}. Note that associations may be identified across slabs (labelled by File number in this illustration), although typically by never more than one superslab on either side.}
    \label{fig:algorithm}
\end{figure*}

\subsection{Design philosophy of \abacus{} merger trees}
\label{sec:philosophy}

Before describing the algorithm in detail, we begin by framing the major concepts and philosophies of our approach to merger trees.

The \summit{} suite is designed to support mock galaxy catalogues based on catalogues and properties of haloes, with rather limited attention to subhaloes.  We seek to provide merger trees to support computation of halo properties such as mass-quantile formation times, to detect major merger events, and to track associations of progenitors and descendents across long stretches of cosmic time.  In principle, one could seek to modulate the properties of central galaxies or quench the growth of infalling satellites based on such associations, such as is done in semi-analytic galaxy formation models.  However, the focus on haloes rather than subhaloes does limit the ability to attempt to model individual satellite galaxies in more physical detail.

Because of this focus on halo scales and associations over longer time, we build the \summit{} merger trees from a moderately dense set of epochs.  We use 33, with a separation in redshift of $\Delta z = 0.05$ at $z\leq0.5$, by $\Delta z = 0.075$ for $0.5<z<1.7$, and somewhat coarser spacing at higher redshift.  This corresponds to a typical time interval of $\Delta t = \Delta \ln a/H(a)\approx (0.04--0.05)/H(a)$, where $a$ is the cosmological scale factor and $1/H(a)$ is the Hubble time.  The choice of 33 is lower than the 60--200 that is commonly adopted in other simulation suites \citep[e.g.][]{Srisawat2013,Wang2016}, but for halo applications we find it to be sufficient (as will be tested in Section~\ref{sec:cadence}).  We stress that the typical halo crossing time (diameter divided by circular velocity) at the L1 radius, given its associated enclosed overdensity, is roughly $t_{{\rm cross}}\approx0.2/H(a)$.  Hence, \summit{} provide 4-5 output epochs per halo crossing time, giving an intuitive reason as to why our merger-tree results are adequately converged against cadence.  We caution that associating importance to halo events on time scales shorter than a crossing time is probably not physically motivated, e.g., the intrahalo medium cannot equilibrate faster than the sound crossing time.  Higher cadences are motivated by the desire to track subhalo mergers, which utilise a higher density threshold and hence smaller dynamical time.  Of course, limiting the number of output epochs is also important to economise on the data volume of the simulations, which was 2 PB even for this adopted set.

Our methodology of building merger trees is based on match particle membership, utilising the fact that \summit{} did track and output unique particle identification numbers.  This allows for a direct association, without the need to use dynamical predictions of halo orbits.  This methodology is common to several algorithms in the literature and has been shown to result in merger trees with greater stability \citep[see, e.g. ][for details]{Srisawat2013}. In this sense, the algorithm presented in Section~\ref{sec:algorithm} is readily generalisable to any halo catalogues that record the IDs and densities of the particle sets defining haloes and is not limited to use in conjunction with \compaso{}.  

One application of the association of haloes between epochs is to detect inconsistencies between single-epoch halo catalogues, in particular in regard to the delicate and inexact problem of deblending of nearby density peaks.  Different halo finders will produce different results on such situations, and there is no perfect answer.  As described in Section~\ref{sec:cleaning}, we use the merger trees to investigate such issues in CompaSO, resulting in a data product of cleaned halo catalogues at a series of epochs.  We recommend that these cleaned catalogues are an improved starting point for the construction of mock galaxy catalogues such as from the halo occupation distribution model. 

\subsection{Merger tree algorithm}
\label{sec:algorithm}

We now describe in detail the algorithm employed for constructing halo merger trees in \summit{}. Note that a `halo' in this context refers to L1 haloes as defined in Section~\ref{sec:compaso}.
% \compaso{} identifies L0, L1, and L2 haloes at each of the 12 primary redshift outputs in \summit{}, as well as at 21 secondary redshifts. This relatively high cadence of outputs is necessary for constructing halo merger trees, the procedure for which is the main focus in this section. Note that a `halo' in this context refers to L1 haloes as defined in Section~\ref{sec:compaso}. We do not track subhaloes in \summit{}. 

To be able to accurately match haloes across output times, we make use of subsampled particles that are output alongside the halo catalogues themselves. In \summit{}, the subsampled particle list is split into 3\% and 7\% (yielding a total of 10\%) sets, which, respectively, are referred to as subsamples ``A'' and ``B''. These subsamples are selected based on a hash of their (unique) particle ID number, and are consistent across redshifts. Each particle is assigned a 64-bit integer, {\tt PID}, which stores both the ID number as well as its kernel density. The ID is simply the $(i,j,k)$ index location of the particle in the initial grid; these numbers are stored as the lower three 16-bit integers in {\tt PID}. The kernel density is stored as the square root of the density (in units of the cosmic density) in bits 1..12 of the upper 16-bit integer in {\tt PID}. The sets of particles associated with L1 haloes are stored contiguously in the subsampled {\tt PID} files; for the purpose of merger tree construction, we only consider those particles that are marked as being part of a halo (i.e., excluding subsampled particles in the ``field''). 

With the list of L1 haloes and their corresponding particles at hand, the merger tree construction, in {\it reverse time order}, proceeds as follows:
\begin{enumerate}
    \item Start with the first L1 halo, {\tt halo\_now}, identified at snapshot {\tt i}, and retrieve its centre (defined by the centre-of-mass of its largest L2 halo) and its 10\% particle subsample.
    \item Identify the list of all haloes in the {\it two preceding} snapshots, {\tt i-1} and {\tt i-2}, that could have {\it plausibly} been associated with {\tt halo\_now}. A plausible association is defined as a halo that is located at most at a distance of $r_{{\rm max}}$ from the centre of {\tt halo\_now}. Throughout, we assume $r_{{\rm max}}=4$ Mpc$/h$.
    \item From the list of plausible associations, we identify {\it candidate associations} as those objects that share a non-zero fraction of their unique particle IDs with {\tt halo\_now}. The fraction of particles donated by these candidate associations to {\tt halo\_now} is labelled as $f_{{\rm donate}}$. We also record the fraction of subsampled particles in {\tt halo\_now}, {\it weighted by kernel density}, that is donated to it by its candidate associations in snapshots {\tt i-1} and {\tt i-2}. We denote these values by $f_{{\rm match}}$.
    \item We then mark a candidate association as a {\tt Progenitor} if $f_{{\rm donate}}\geq f_{{\rm thresh}}$. The remaining candidate associations are discarded. We assume $f_{{\rm thresh}}=0.5$ throughout. 
    \item We identify the {\tt MainProgenitor} as the halo association at snapshot {\tt i-1} that contributes the largest $f_{{\rm match}}$. We further record {\tt MainProgenitorPrec} (`preceding'), which is the {\tt MainProgenitor} identified in snapshot {\tt i-2}. While we keep track of the full {\tt Progenitor} list from snapshot {\tt i-1}, we only track the value of {\tt MainProgenitorPrec} from snapshot {\tt i-2}.
    \item Repeat steps (i)-(v) for the remaining haloes at snapshot {\tt i}. 
    \item Repeat steps (i)-(vi) for all subsequent triples of output times in a given simulation.
\end{enumerate}
A cartoon illustration of this halo association procedure across any two \summit{} snapshots is shown in Figure~\ref{fig:algorithm}. For any given halo, the joint list of its {\tt Progenitors} and {\tt MainProgenitors} across all output times defines its merger tree. To conveniently pinpoint a halo in the simulation, we define a unique halo identifier, {\tt HaloIndex}, defined as:
\begin{multline*}
    {\tt HaloIndex}=1e12*{\tt FullStepNumber}+1e9*{\tt SuperslabNumber}\\+{\tt IndexInSuperslabFile},
\end{multline*}
where {\tt FullStepNumber} is the number of the timestep at which the halo catalogue has been output, and {\tt IndexInSuperslabFile} is the array index of the halo in the corresponding halo catalogue file. The {\tt Progenitor} and {\tt MainProgenitor} lists for a given halo are then simply collections of {\tt HaloIndex} values appended over multiple output times.

As described in the methodology above, haloes identified at any given output time are matched across {\it two} preceding snapshots simultaneously. The primary advantage of this is that it allows us to keep track of fly-bys between pairs of haloes that might occur between one output time and the next. For example, consider the case of two haloes that were previously separate objects at snapshot {\tt i-2}, ``merge'' momentarily at snapshot {\tt i-1}, and then again split away at snapshot {\tt i}. In this instance, for at least one halo in this pair, 
\begin{multline*}
    {\tt MainProgenitor} \left({\tt MainProgenitor}\left({\tt halo\ at\ snapshot\ i}\right)\right) \\ \neq {\tt MainProgenitorPrec}\left({\tt halo\ at\ snapshot\ i}\right)
\end{multline*}
In other words, tracking {\tt MainProgenitorPrec} allows us to ``switch'' to the correct branch of the merger tree on the occasions where a transient fly-by occurs.

One subtlety that arises in our method pertains to step (ii) in the algorithm, where the list of plausible halo associations is narrowed down by first filtering out haloes that are located more than a distance $r_{{\rm max}}$ at the preceding output time. Since haloes are split across multiple files in superslabs (see Section~\ref{sec:compaso}), it may be the case (particularly for haloes located close to the boundary of a superslab file) that several plausible halo associations may found in other superslabs, although typically by never more than one superslab on either side (in the \textsf{Base} simulations, each superslab has a width of around 60 Mpc$/h$). For example, a halo located in superslab file 1 at snapshot {\tt i} may have candidate association haloes (and, therefore, progenitors) located in superslab files 0, 1, or 2 at snapshot {\tt i-1} (see Figure~\ref{fig:algorithm}). For this reason, at any given step of our merger tree algorithm, we are required to simultaneously hold in memory halo and {\tt PID} catalogues for 9 superslabs at any given time: 3 per snapshot, and 3 snapshots in total ({\tt i}, {\tt i-1}, {\tt i-2}).

Finally, in some rare circumstances, we identify associations where one halo is marked as the {\tt MainProgenitor} of another, but does not also appear in the {\tt Progenitor} list of this halo. While this may at first seem logically inconsistent, such a situation can occur when one halo overwhelmingly contributes the largest $f_{{\rm match}}$, but does not actually donate the majority of its particles to the same halo (i.e., $f_{{\rm donate}}<f_{{\rm thresh}}$). This happens in the case of `splits': when a low-mass `halo' that once belonged to a larger aggregate unit breaks off at a later time, either due to the object's orbital trajectory, which ejects it from its parent halo, or due to a deblending decision made by \compaso{}. In this sense, a split may be physical (in the former case) or numerical (in the latter). In the outputs produced by the merger tree calculation, these incidences are marked with a boolean flag, {\tt IsPotentialSplit}, which is set to {\tt 1} when the condition above is met. Note that these cases are infrequent: occurring in $\sim2\%$ of 150-particle haloes, and in $0.001\%$ of 15,000-particle haloes (see Section 3.4 and Table 2 of \citealt{Hadzhiyska2021}).

\subsection{Code optimisation routines}
\label{sec:optimisation}

Our code for building halo merger trees from \summit{} has been developed to be used as a post-processing tool, and is written in Python (although some of the packages it uses are Python wrappers around C code). The task of determining halo associations for 100s of millions of haloes across multiple output times for each simulation in the \summit{} suite is demanding: for this reason, we employ a few optimisation routines that dramatically improved the performance of our code:
\begin{itemize}
    \item Step (ii) of the algorithm, which first performs a neighbour search around the halo of interest to narrow down the list of all plausible halo associations from the previous output time significantly reduces the number of objects whose particle list have to be cross-matched. The neighbour search is performed using a multi-threaded tree search, for which we use the {\tt scipy} package's {\tt cKDTree} implementation \citep{Scipy2020}.
    \item Steps (iii)-(v), which are performed on a per-halo basis, is accelerated by a factor of $\sim15\times$ using the {\tt Numba} just-in-time compiler \citep{Numba}, which translates Python functions to optimised machine code at runtime.
    \item The loop over all haloes in a single step is a classic example of an `embarrassingly parallel' problem, whose speed scales linearly with the number of CPU processes employed for the task. We address this using the {\tt joblib} library\footnote{\url{https://joblib.readthedocs.io/}}, which uses Python's {\tt multiprocessing} functionality to achieve such a speedup. On Rhea\footnote{\url{https://www.olcf.ornl.gov/olcf-resources/compute-systems/rhea/}}, we use 16 cores for processing \summit{} merger trees, which yields another $16\times$ speedup. 
\end{itemize}
Finally, we note that that merger trees are not constructed for every single halo in the \summit{} suite, but only for those in which particle tracking can be carried out reliably. We determine this to be the case when a halo contains at least 50 particles in total, with at least 5 subsampled particles available for tracking. The 50-particle limit corresponds to a lower limit of $1.05\times10^{11}\ {\rm M}_\odot/h$ in halo mass at the \textsf{Base} mass resolution.

\section{Cadence of outputs for building merger trees}
\label{sec:cadence}

In this section, we present statistics of merger trees constructed from our algorithm for different cadences in the number of snapshots used to build halo trees. The aim of this section is to explore the influence of our choice of 33 intervals (c.f. Section~\ref{sec:philosophy}) for typical \summit{} simulations.

To this end, we have run a dedicated simulation set which we refer to as \cadence{}. This simulation has a box size of  296 Mpc$/h$ with $1024^3$ particles, resulting in an effective mass resolution that is identical to \summit{} {\sf Base} resolution. We output 186 snapshots separated in cosmological scale factor by $\Delta \ln a \approx 0.007$, corresponding to roughly 6$\times$ the density of outputs of a typical \summit{} simulation. To perform our tests, we subsequently build trees using all 186 outputs (highest cadence), 64 outputs ($\approx 3\times$ coarser, medium cadence), and 33 outputs ($\approx 6\times$ coarser, Summit-like cadence). 

We begin by computing perhaps the simplest diagnostic tool for assessing halo merger trees which is their mass accretion histories i.e., quantifying the mass growth of a halo as a function of redshift. The most common way of expressing this information is to traverse the main progenitor branch of individual halo merger trees, and to parameterise the  accretion history as simply the mass evolution of the main progenitor.

\begin{figure}
    \centering
    \includegraphics[width=\columnwidth]{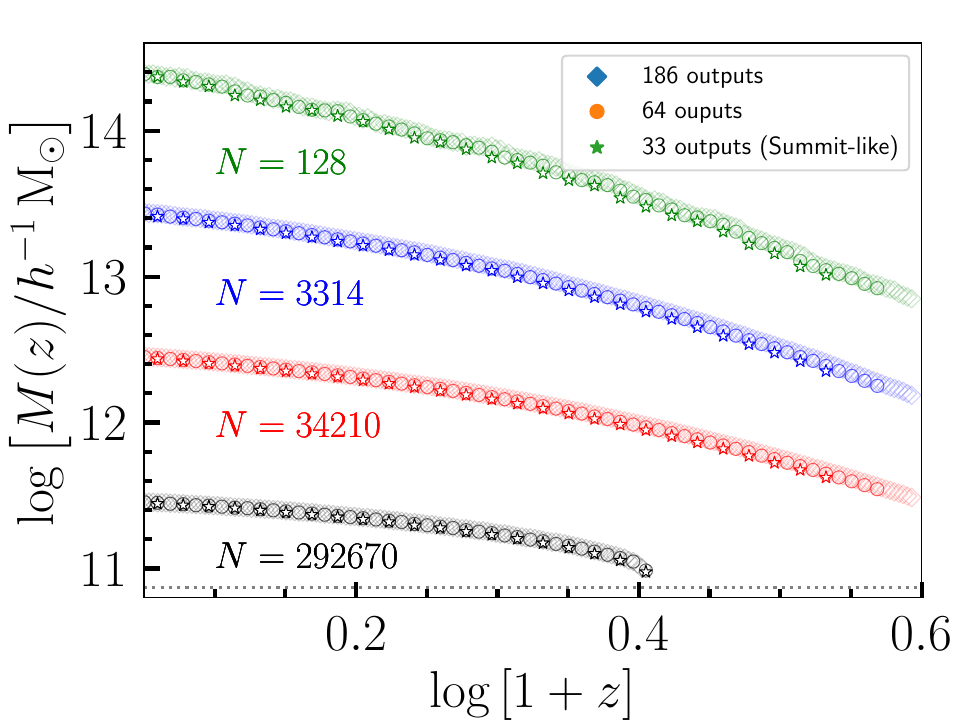}
    \caption{Median mass accretion histories of haloes identified at $z=0.1$ in the \cadence{} simulation from merger trees built using the fiducial spacing of output times used in \summit{} (stars, 33 outputs), and with $\sim 2\times$ and $\sim 6\times$ denser outputs (circles and diamonds, respectively). The grey horizontal dotted line marks the 50-particle limit, which corresponds to the smallest haloes for which merger tree associations are tracked. In each case, we follow haloes along the {\tt MainProgenitor} branch of their merger tree. We find excellent agreement in the accretion histories irrespective of the density of outputs used to build the merger tree.}
    \label{fig:cadence_mah}
\end{figure}

\begin{figure}
    \centering
    \includegraphics[width=\columnwidth]{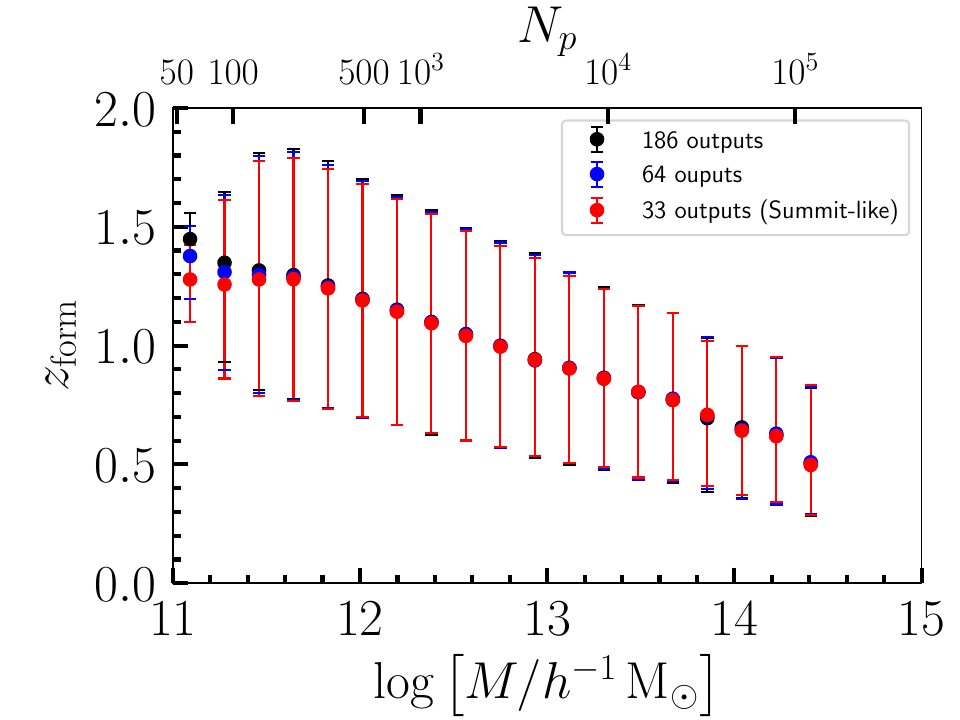}
    \caption{The redshift of formation, $z_{{\rm form}}$, of haloes in the \cadence{} simulation for objects identified at $z=0.1$. The epoch of formation is defined as the redshift at which the halo attains 50\% of its final mass. The circles mark the median $z_{{\rm form}}$, while the error bars indicate the 16$^{{\rm th}}$-84$^{{\rm th}}$ percentile of this distribution. Both the median and the scatter in the measurement of formation time are near identical irrespective of whether 186 outputs or 33 outputs are used to construct the merger tree.}
    \label{fig:cadence_zform}
\end{figure}

Figure~\ref{fig:cadence_mah} shows the median mass accretion histories associated with haloes in bins of mass centred at $\log\left[ M_0 /{\rm M}_\odot/h \right] = \left[11.5, 12.5, 13.5, 14.5\right]$, each with a width of $\pm 0.2$ dex. Here, $M_0$ is the mass at $z=0.1$. The results for trees with 186 outputs, 64 outputs, and 33 outputs, respectively, are represented by diamonds, circles, and stars. It is clear from this figure that mass accretion histories are recovered equally well irrespective of the temporal density of the trees, with the agreement at $z\lesssim1.5$ in particular being excellent. At higher redshift, there is a noticeable deviation in the agreement between the Summit-like accretion histories and the highest cadence trees, particularly in the most massive bin, where we average over only 128 haloes. The lower cadence merger trees tend to slightly underestimate the mass of the main progenitor at high redshift, which could potentially be due to the algorithm choosing a different main progenitor branch when there are two similar mass progenitors from which to choose. This can happen when a halo's {\tt MainProgenitor} in the previous timestep is identified as a singular entity in fine-grained trees (which looks over shorter time intervals), but may be split between two objects when looking back over the longer time interval in the coarser trees. However, the effect is small: in the most massive bin, the typical agreement (across all redshifts) in the mass of the {\tt MainProgenitor} is of the order of $\sim4\%$ between the highest cadence and the Summit-like cadence trees. In lower mass bins, the agreement improves to within 1\%.

Figure~\ref{fig:cadence_zform} shows a summary statistic that can be used to quantify the accretion histories of haloes, the so-called redshift of formation, $z_{{\rm form}}$, as a function of halo mass. This parameter can be used to augment mass-only HOD models to e.g. incorporate the dependence of galaxy occupation and/or properties on halo age \citep[e.g.][]{Hearin2013}. We measure $z_{{\rm form}}$ by traversing the main progenitor branch of the merger tree and identifying the formation redshift as the epoch at which the halo achieves 50\% its final-day mass. 

\begin{figure}
    \centering
    \includegraphics[width=\columnwidth]{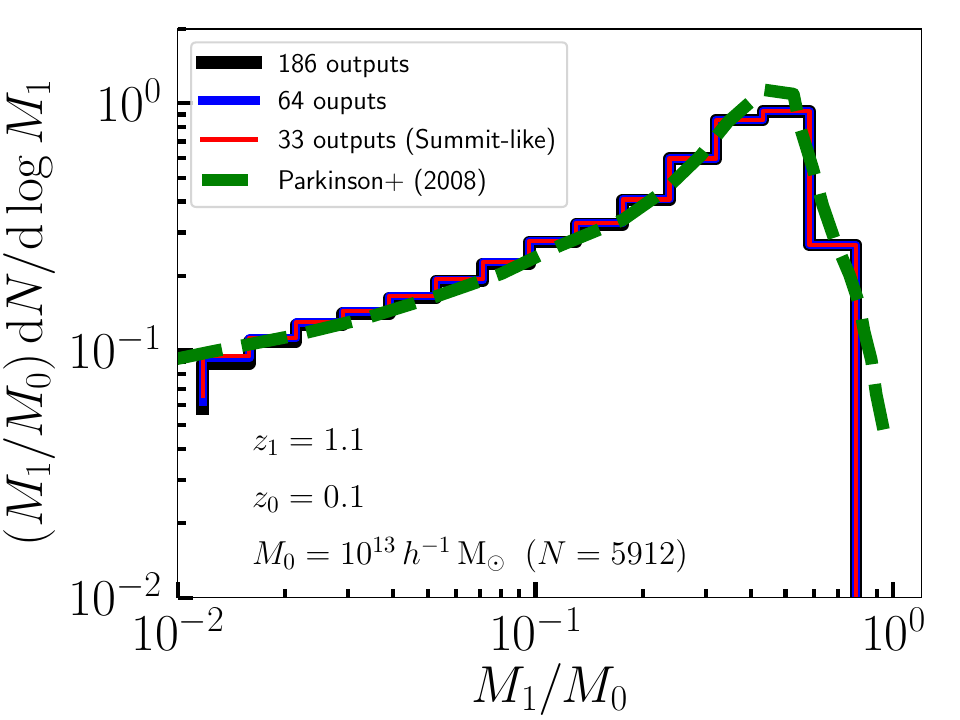}
    \caption{The fraction of mass contained in haloes of mass $M_1$ at $z=1.1$ that merge into haloes of mass $10^{13}\,h^{-1}\,{\rm M}_\odot$ by $z=0.1$ measured in the \cadence{} simulation. The so-called `conditional mass function' is nearly identically irrespective of how many outputs are used to build the merger tree. The dashed green curve represents the conditional mass function predicted by the algorithm of \citet{Parkinson2008}, which is based on a modified version of  extended Press-Schechter theory.}
    \label{fig:cadence_cmf}
\end{figure}

The mass dependence of halo formation time is shown in black, blue, and red, respectively, for the high, medium, and Summit-like cadence merger trees. The symbols represent the median value of $z_{{\rm form}}$, while the error bars span the 16$^{{\rm th}}$-84$^{\rm th}$ percentile of the distribution. As expected, there is a negative correlation between $z_{{\rm form}}$ and halo mass, indicating that low-mass haloes are older than high-mass haloes, which is the standard expectation from hierarchical structure formation. There is excellent agreement between all three versions of the merger tree calculation, both in the median and in the scatter. Encouragingly, the consistency between the three calculations is preserved all the way down to the 50-particle limit. This test demonstrates that for estimating the formation epoch of haloes, which is a common use case for merger trees generally, the Summit-like output cadence of 33 snapshots is sufficient. 

Finally, we investigate how the density of outputs used to construct a merger tree affects the identification of progenitors beyond the primary branch. Figure~\ref{fig:cadence_cmf} shows the {\it conditional} mass function, which represents the fraction of mass contained in progenitor haloes of mass $M_1$ at $z=1.1$ that merge into haloes of mass $10^{13}\,{\rm M}_\odot/h$ by $z=0.1$. We construct this quantity by stepping through every time slice in the merger tree, and recording the progenitors (and progenitors of progenitors) whose final descendant points to a $10^{13}\,{\rm M}_\odot/h$ halo at $z=0.1$ (of which there are 5912 examples in the \cadence{} box).

The measurements from the high, medium, and Summit-like cadence merger trees, respectively, are represented by the black, blue, and red histograms. We find that the conditional mass function is reproduced near identically in all three cases, demonstrating the fact the complete progenitor histories of haloes are recovered even using the coarser spacing of snapshots used to build merger trees in \summit{}. The three versions of the trees agree down to progenitor mass ratios of $M_1/M_0=10^{-2}$, corresponding to haloes of mass $10^{11}\,{\rm M}_\odot/h$ at $z=1.1$, which contribute $\approx 10\%$ of the final mass of the descendant halo. The dashed green line shows the conditional mass function predicted by the Monte Carlo algorithm of \cite{Parkinson2008}, which is based on an extension of extended Press-Schechter theory, calibrated to reproduce the results of the Millennium simulation \citep{Springel2005}. Reassuringly, we find very good agreement between the conditional mass function predicted by this model, to that extracted from the merger tree algorithm presented in this work.

The results in this section demonstrate that the snapshot spacing we have chosen for constructing  trees in \summit{} is sufficient for halo-level merger tree statistics, particularly for the kinds of metrics that would typically be used in building mock galaxy catalogues. In the subsequent section, we apply our algorithm to the \summit{} suite, and present some initial applications of the resulting data products.

\section{Applications of the merger tree}
\label{sec:applications}

In the following subsections, we present some initial applications of the halo merger trees constructed from 
\summit{} using the algorithm that has been described in Section~\ref{sec:algorithm}. While merger tree associations have been constructed for all simulations in the \summit{} suite, here we report results from the \highbase{} and \high{} simulations only (both at the \cph{} cosmology). 

\subsection{Halo mass accretion histories}
\label{sec:mass_accretion}

\begin{figure}
    \centering
    \includegraphics[width=\columnwidth]{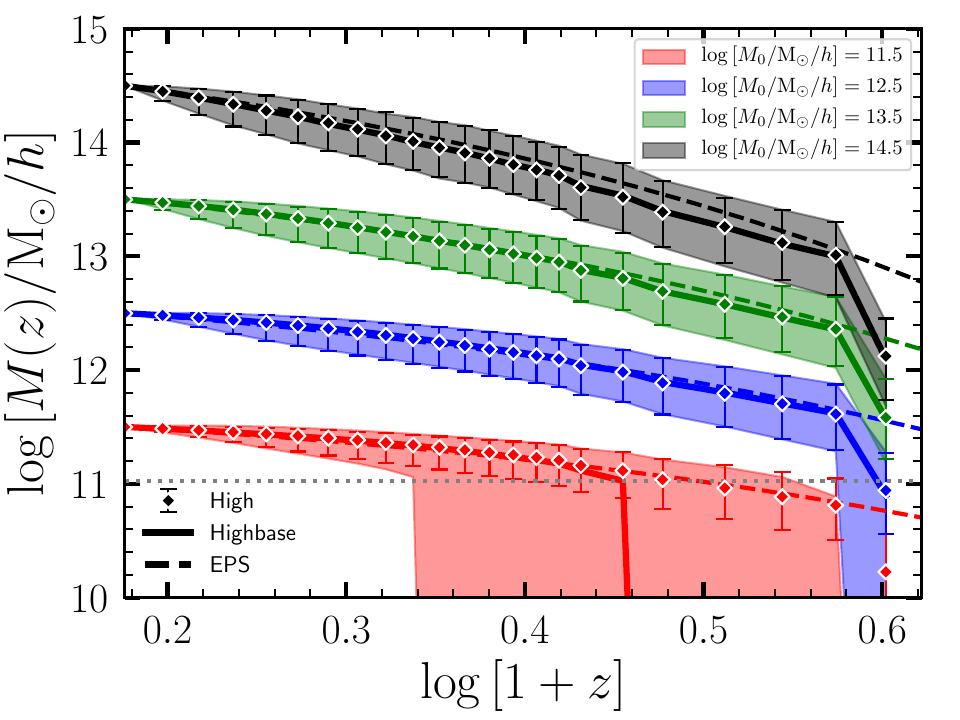}
    \caption{Median mass accretion histories of haloes identified at $z=0.5$ in the \cph{} cosmology at the \highbase{} (solid lines) and \high{} (symbols) resolution, where the latter has $6\times$ better mass resolution then the former. The mass bins are centred at $\log\left[ M_0 /{\rm M}_\odot/h \right] = \left[11.5, 12.5, 13.5, 14.5\right]$, each with a bin width of $\pm 0.2$ dex. The shaded bands (error bars) encompass the $16^{{\rm th}}$ and $84^{{\rm th}}$ percentiles of the scatter in the accretion histories for \highbase{} (\high{}). The dashed curves show predictions for the accretion histories using the semi-analytic model described in \citet{Correa2015}. The grey horizontal dotted line marks the 50-particle limit in \highbase{}, which corresponds to the smallest haloes for which merger tree associations are tracked. }
    \label{fig:mah_convergence}
\end{figure}

In Figure~\ref{fig:mah_convergence}, we present the median mass accretion histories of haloes identified at $z=0.5$ in the \cph{} \highbase{} (solid lines and shaded regions) and \high{} (symbols with error bars) simulations. Recall that the latter has $6\times$ better mass resolution in the same 1 Gpc$/h$ box. The symbols/lines represent the median accretion history while the error bars/shaded regions encompass the $16^{{\rm th}}$ and $84^{{\rm th}}$ percentiles of the scatter around the mean. Different colours represent accretion histories from different halo mass ranges, where we have set the bin centres to be at $\log\left[ M_0 /{\rm M}_\odot/h \right] = \left[11.5, 12.5, 13.5, 14.5\right]$, each with a width of $\pm 0.2$ dex. In \highbase{}, this yields 11574539, 1251068, 96822, and 1501 haloes, respectively. 

We notice immediately that the two simulations show excellent agreement in the regions where they overlap, despite the factor of 6 difference in mass resolution. This provides a simple, yet reassuring, test of numerical convergence in two of our \summit{} simulations, at least from the perspective of this diagnostic for halo merger trees. This is seen most clearly in the lowest mass bin, where we see a sharp break in the accretion histories extracted from \highbase{} below the 50-particle limit, which is indicated using the horizontal dashed line in grey. Beyond this limit, we see a smooth transition into the regime that is now resolved in the \high{} simulation.

For comparison with expected trends, Figure~\ref{fig:mah_convergence} also shows predictions for the mass accretion history of haloes using the semi-analytic model developed by \cite{Correa2015}, which is itself inspired by the extended Press-Schechter formalism. In general, there is good agreement between this model and our measurements from the simulations, although there are slight differences in the most massive mass bin. This is perhaps unexpected, as the model by \cite{Correa2015} has been calibrated against a different set of numerical simulations. Furthermore, the comparison in the largest mass bin may be compromised by the relatively small number of objects that fall under this mass range (624). 

\subsection{Cleaning \compaso{} catalogues}
\label{sec:cleaning}

As we have discussed in the Introduction and at the end of Section~\ref{sec:algorithm}, there is often ambiguity when it comes to reliably tracking the halo population in a cosmological simulation. A wide variety of processes including fly-bys, partial mergers, splits, or simply limited numerical resolution can undermine the robustness of halo properties output at any given time. Much like in the construction of the trees themselves, there are a variety of ways in which these fractures in the merger tree may be addressed. Examples include algorithms that try to `patch' trees across gaps in a halo's history by inserting additional haloes into the catalogue that guarantee smoother mass evolution \citep[e.g.][]{Tweed2009,Knebe2010,Behroozi2013b,Jiang2014}. Halo splits (into multiple descendants) may well be treated as physically allowed scenarios, such as when two haloes that are in reality distinct dynamically but have at some point been treated as a single object by the halo finder. Characterising the split objects appropriately requires careful consideration of their energy and phase space properties; recent works that describe the treatment of these splits include those by \cite{Han2018} and \cite{Roper2020}.

In \summit{}, another instance when pathologies in the halo catalogue may arise is when the \compaso{} algorithm has been overly aggressive in deblending single haloes into two or more components. This may happen due to the strict spherical overdensity criterion in \compaso{} that draws hard edges in the ellpisoidal halo particle distribution, or indeed \compaso{}'s eligibility condition that is adept at finding new halo nucleation points on the outskirts of haloes \citep[see][for details]{Hadzhiyska2021}. A general effect of these processes is to detach particles from haloes, leading to a reduction in their mass relative to the time when a potential galaxy hosted by this halo may have formed. A model like the HOD, which determines the galaxy occupancy of haloes based on their mass at any given time may therefore incorrectly predict the galaxy content of such objects. 

To overcome these issues in \summit{}, we take a conservative approach and simply `clean' the \compaso{} halo catalogues of objects that may have been compromised by the processes listed above. One way to do this is to simply flag a halo that at some epoch passes through (or flys-by) another, sharing some of its particles in the process, before re-emerging as two separate objects at some later time. We then ``merge'' the particles of the two haloes into a singular entity and, for all subsequent output times, assume that this agglomerated object remains as a persistent entity (unless it itself undergoes a future fly-by/splashback event). In general, this method of cleaning halo catalogues in post-processing removes $\sim 1-5\%$ of objects (see Table~\ref{tab:clean_table}), predominantly composed of a population of low-mass haloes floating around the boundaries of more massive haloes, and appends their particle lists to these  neighbours. Note that the same technique will also result in re-merging overly deblended haloes.

The task of identifying and re-merging the offending haloes is simplified with the help of halo merger trees. In detail, our cleaning procedure proceeds as follows:
\begin{figure*}
    \centering
    \includegraphics[width=\textwidth]{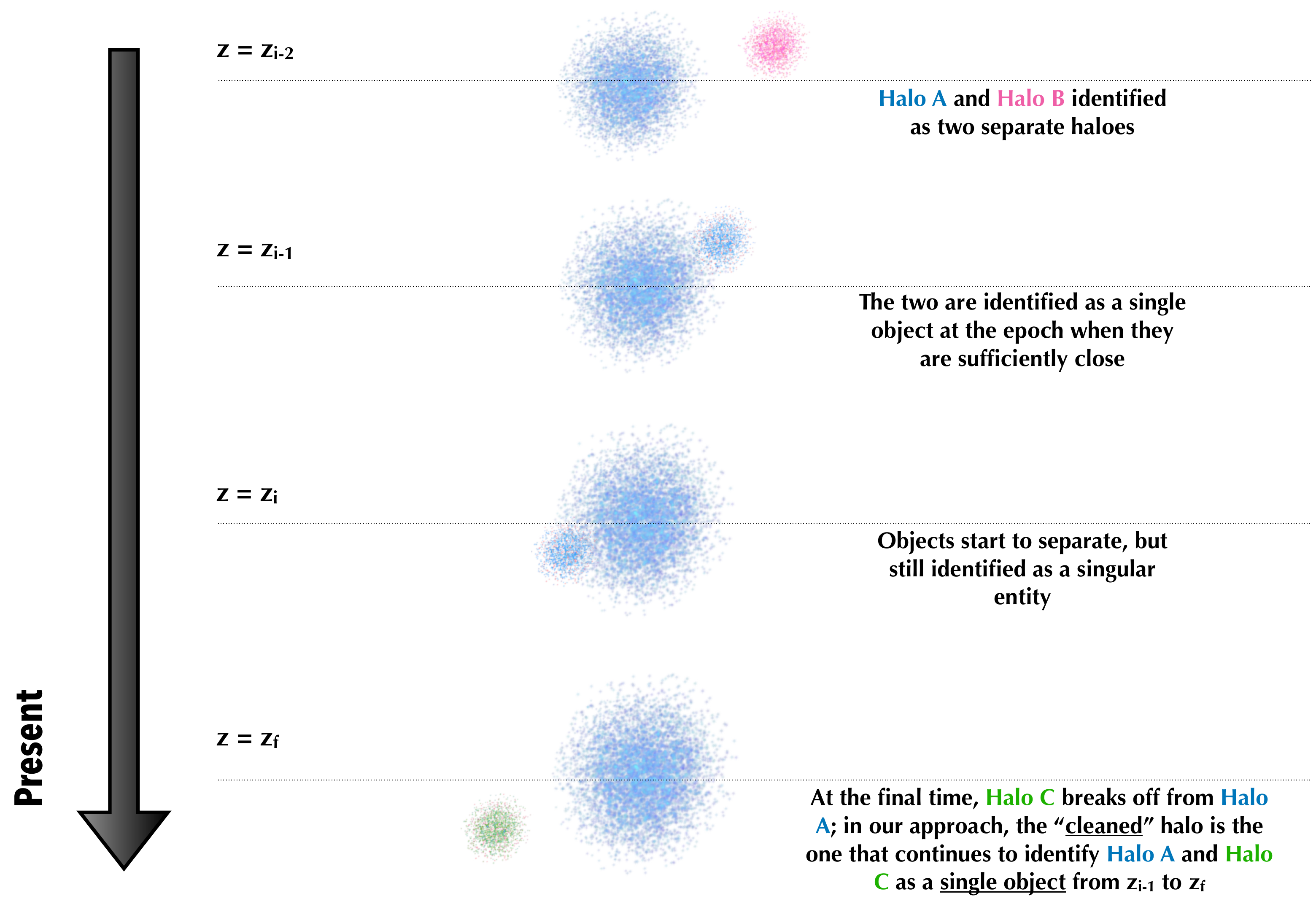}
    \caption{A schematic diagram showing the effect of cleaning as described in Section~\ref{sec:cleaning}. At some early epoch, $z=z_{i-2}$, objects marked as Halo A (blue) and Halo B (pink) are identified as two distinct haloes. They subsequently interact and share particles such that they are then labelled as belonging to the same aggregate halo. At some later time, $z_f$, particles that once belonged to Halo B (or perhaps a subset of particles from B and some from A) may break off from this aggregate body, and be identified as some separate set C (green). In some halo catalogues, the green particles may be identified as a separate halo in its own right. In our ``cleaning'' approach, we continue to treat the green and blue particles as one object.}
    \label{fig:cleaningflow}
\end{figure*}
\begin{figure}
    \centering
    \includegraphics[width=\columnwidth]{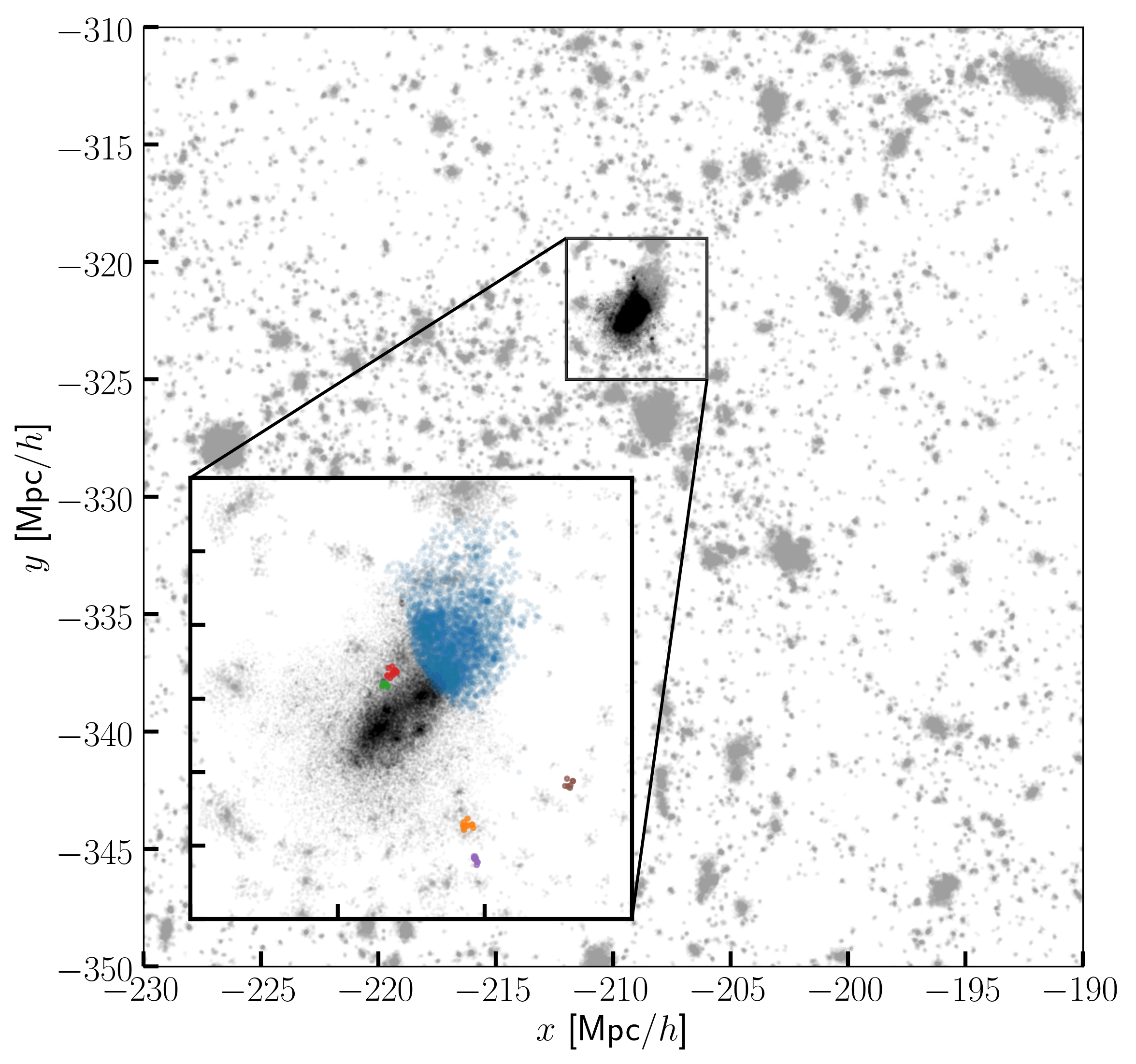}
    \caption{A visual depiction of the cleaning procedure outlined in Section~\ref{sec:cleaning}. The background panel shows a (40 Mpc$/h$)$^{3}$ region where the most massive halo ($M_{{\rm halo}}=10^{15} {\rm M}_\odot/h$) is represented in black. The remaining haloes are displayed in grey where, for clarity, we only show haloes more massive than $10^{11}{\rm M}_\odot/h$. In each case, we only display the 10\% subsampled particles associated with each halo. The inset panel zooms into region of size $6\times6\times40$ $\left({\rm Mpc}/h\right)^{3}$, in which we have now highlighted the haloes (6 in total, represented by the different colours) that are flagged by our cleaning algorithm and merged with the $10^{15} {\rm M}_\odot/h$ halo. While most of the merged objects are of relatively low mass ($\sim10^{11} {\rm M}_\odot/h$), the halo represented by the blue points is a rather massive object, with $M_{{\rm halo}}=1.02\times10^{14} {\rm M}_\odot/h$. This is a clear example where the \compaso{} algorithm has deblended two haloes based on the density criteria described in Section~\ref{sec:compaso}. The net effect of merging these haloes with the neighbouring cluster is to increase the mass of the central host by roughly 10\%.}
    \label{fig:clean_example}
\end{figure}
\begin{enumerate}
    \item For a halo identified at redshift $z_i$, record its \compaso{} mass, $M_{{\rm halo}}\left(z_i\right)$.
    \item Traverse its merger tree along the {\tt MainProgenitor} branch, and identify the epoch, $z_{{\rm max}}$, at which it achieves its maximum mass, $M_{{\rm halo}}\left( z_{{\rm max}}\right)$.
    \item If the ratio $\frac{M_{{\rm halo}}\left( z_{{\rm max}}\right)}{M_{{\rm halo}}\left(z_i\right)} > \kappa$, flag this halo for cleaning.
    \item At redshift $z_{{\rm max}}$, find the most massive halo whose {\tt HaloIndex} is equal to the {\tt MainProgenitor} of the flagged halo at this time. Once this entity has been found, add the mass of the flagged halo to the mass of its massive neighbour.
    \item The aggregated halo now remains merged for all subsequent outputs. The flagged halo is eliminated from the halo catalogue at each of the subsequent outputs. The particle list of the aggregate halo is the union of the particles of the original massive halo and {\it all} haloes that have been merged with it.
\end{enumerate}
We display an illustration of the effect of cleaning in Figure~\ref{fig:cleaningflow}. Here, we start with having two sets of particles labelled as distinct haloes A (blue) and B (pink) at some early epoch, $z_{i-2}$. As these objects come together and interact, at some point they will share enough particles that the halo finder will label them as belonging to the same aggregate halo (between $z_{i-1}$ and $z_i$). As the erstwhile Halo B emerges out of this aggregate at $z_f$, it may now be composed of some combination of particles from A and B: we label these in green. Different halo catalogues will treat the green particles differently, typically identifying it as some new Halo C, whose mass and other properties will be the outcome of the complex interaction between Halo A and Halo B. In the cleaning procedure we have described in this section, the particles in the green and blue sets continue to be treated as a single, aggregate halo (i.e. all the way from $z_{i-1}$ to $z_f$. In other words, the particle list in green is appended to the particle list in blue.
A concrete example of this procedure applied to the \cph{} \highbase{} simulation is shown in Figure~\ref{fig:clean_example}, where in the different colours we highlight some of the haloes that are ``merged'' onto some nearby massive neighbour following the cleaning strategy.

In very rare instances ($\sim0.1\%$ of the time), a halo may be flagged for cleaning but no distinct host halo is identified in step (iv) of the procedure outlined above. This is typically the case for two initially separate objects that just come into contact for the first time, where \compaso{} assigns more than half the particles associated with the smaller halo to its more massive neighbour. The larger halo will not be marked as the {\tt MainProgenitor} of the smaller one, but may appear in the complete list of {\tt Progenitors}. We then merge the smaller object with this halo. If no match is found even in the {\tt Progenitors} list, we then simply merge the flagged halo with the nearest object that contains at least as many particles as the difference in the particle number between the peak and present-day mass of the flagged halo. 

 The method outlined in this subsection identifies objects whose present-day mass may be under-counted due to fly-bys, splits, partial merging etc. This algorithm preferentially flags (and removes) low-mass objects, and append their particle lists to heavier, neighbouring haloes that they were once attached to. Note that for objects with ``well-behaved'', smooth merger histories, $M_{{\rm halo}}\left(z_i\right) = M_{{\rm halo}}\left( z_{{\rm max}}\right)$, so they will not be flagged by this procedure. Our method has a single free parameter, $\kappa$, for which our recommended choice is $\kappa=2.0$. Based on our tests, however, we find that the properties of the cleaned halo catalogues do not change significantly for small variations in $\kappa$. In the following subsections, we present comparisons of the statistics of the halo population in the cleaned and default \compaso{} catalogues. 

\subsection{Halo mass functions}
\label{sec:hmf}

\begin{figure}
    \centering
    \includegraphics[width=\columnwidth]{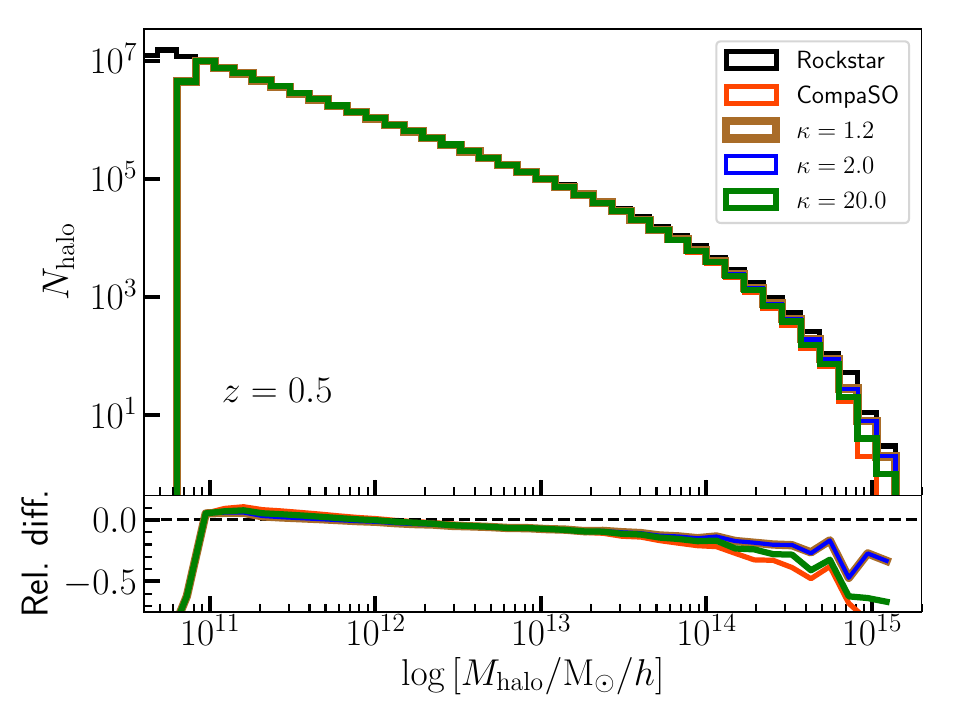}
    \caption{A comparison of the halo mass functions extracted from the \highbase{} \cph{} simulation at $z=0.5$. Halo finding has been performed using both \rockstar{} (black) and \compaso{} (red). The ``cleaned'' versions of the \compaso{} catalogues obtained after assuming cleaning parameters, $\kappa=1.2, 2.0, 20.0$ are shown, respectively in brown, blue, and green (see Section~\ref{sec:cleaning} for details of the cleaning procedure). The standard \compaso{} mass function deviates from \rockstar{} for haloes more massive than $\sim10^{13}\ {\rm M}_\odot/h$, being suppressed relative to it by 50\% at $M_{{\rm halo}}\approx10^{15} \ {\rm M}_\odot/h$. The ``cleaned'' catalogues bring the \compaso{} mass functions into better agreement with \rockstar{}: $\kappa=2.0$, which is our fiducial choice, reduces the discrepancy to 25\%; there is no discernible difference in the $\kappa=1.2$ case. For $\kappa=20.0$, which is the least aggressive cleaning choice, there is only a marginal difference relative to the default \compaso{} catalogue.} 
    \label{fig:hmf_compare}
\end{figure}

Figure~\ref{fig:hmf_compare} compares the halo mass function (measured at $z=0.5$) for the \cph{} \highbase{} simulation for the default (red) and cleaned ($\kappa=1.2$: brown; $\kappa=2.0$: blue; $\kappa=20.0$: green) versions of the \compaso{} catalogues. In black, we also show the halo mass function measured by the \rockstar{} halo finder \citep{Behroozi2013} run on the same simulation. The most significant difference in the way in which \rockstar{} operates is that is a phase-space halo finder, utilising both the positions and velocities of dark matter particles to identify haloes. This provides a very useful, independent method to compare the predictions of \compaso{} and its variations in the form of the cleaned catalogues.

Considering just \rockstar{} and the default \compaso{} (in red) first, we find that there are noticeable differences on both extremities of the halo mass function. In particular, \compaso{} shows a $\sim10\%$ excess in the abundance of $10^{11} {\rm M}_\odot/h$ haloes. On the other hand, at the high-mass end, \compaso{} shows a $\sim50\%$ {\it deficiency} in the abundance of $10^{15} {\rm M}_\odot/h$ haloes. While some of this difference stems inevitably from the different methods used to define haloes in \rockstar{} and \compaso{}, the full extent of the discrepancy cannot be attributed to differences in halo finding. 

\begin{figure*}
    \centering
    \includegraphics[width=\textwidth]{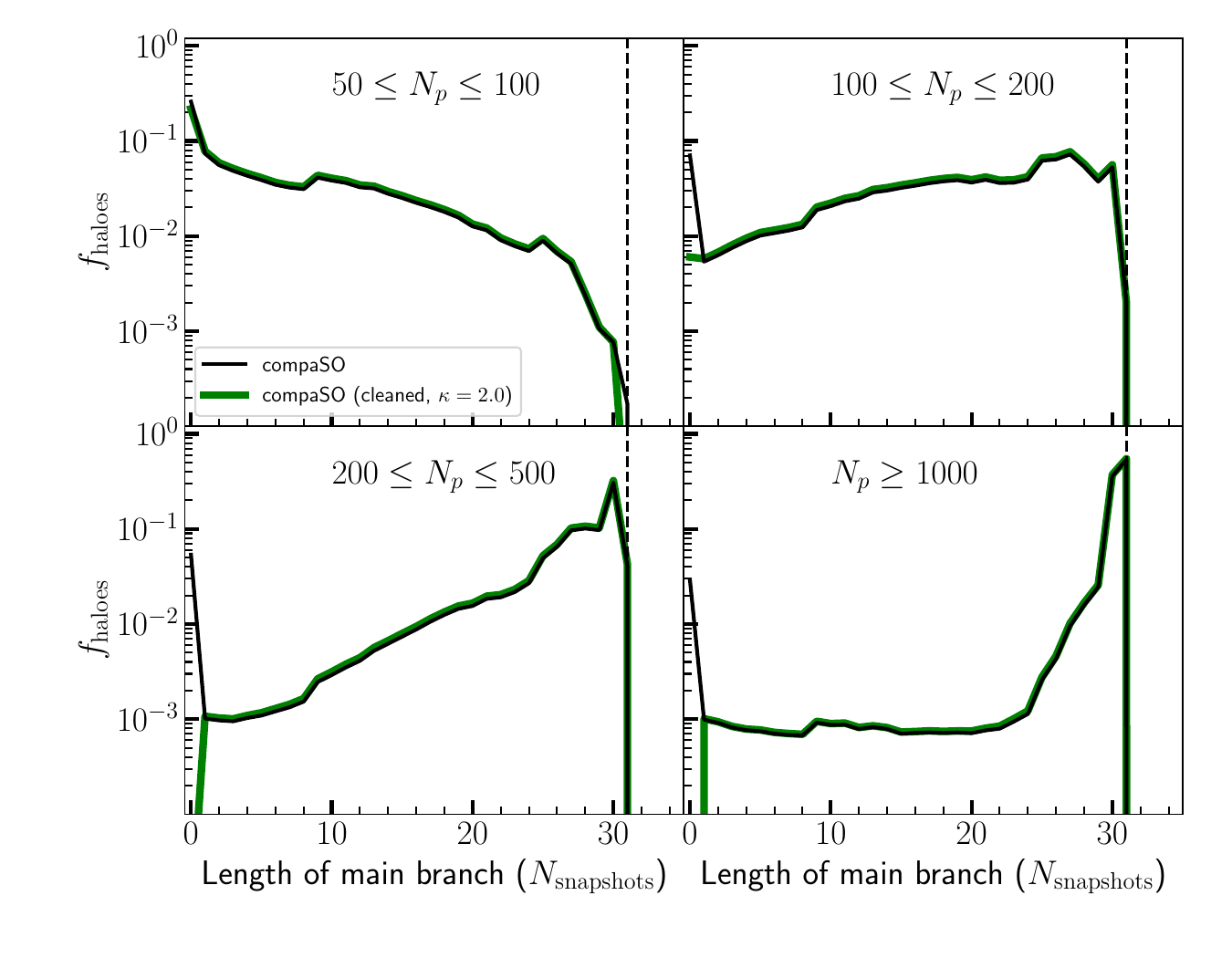}
    \caption{A comparison of the length of the main progenitor branch of haloes extracted from the \highbase{} \cph{} simulation at $z=0.1$. The length of the main progenitor branch is defined as the number of snapshots through which the {\tt MainProgenitor} of any given halo can be tracked back in time. Results for the standard \compaso{} halo catalogue are shown in black, while ``cleaned'' \compaso{} haloes are shown in green. The four panels show the results from four different mass bins selected according to the number of particles in the halo, $N_p$, at $z=0.1$. $f_{{\rm haloes}}$ denies the fraction of haloes in each mass bin that can be tracked for the corresponding number of snapshots. The dashed vertical line marks the total number of snapshots that have been used to create the merger tree (excluding the final output). The standout feature is that in the cleaned version, there is a sharp decrease in the number of haloes that can only be tracked for 0 or 1 snapshots. These correspond predominantly to the population of filtered haloes (i.e. those that have been flagged for cleaning). It is also clear that haloes resolved with more particles are easier to track back in time.}
    \label{fig:length_branch_compare}
\end{figure*}

\begin{table}
    \centering
    \begin{tabular}{c|c|c|c}
    \hline \hline
      $\log\left[ M_{{\rm halo}}/ {\rm M}_\odot/h\right]$&  & $f_{{\rm merged}}$  & $f_{{\rm inc}}$  \\ 
      & $\kappa$ & 1.2\ \ \ \ 2.0\ \ \ \ 20.0 & 1.2\ \ \ \ 2.0\ \ \ \  20.0 \\
     \hline \
      $11.5\pm0.1$ &  &  0.10\ \ \ \ 0.05\ \ \ \ 0.02 & 5$e$-3\ \ \ \ 4$e$-3\ \ \ \ 4$e$-5 \\
      $12.5\pm0.1$ &  &  0.06\ \ \ \ 0.02\ \ \ \ 6$e$-3 & 0.34\ \ \ \ 0.16\ \ \ \ 0.04 \\
      $13.5\pm0.1$ &  &  0.03\ \ \ \ 0.01\ \ \ \ 8$e$-5 & 0.97\ \ \ \ 0.89\ \ \ \ 0.81 \\
      $14.5\pm0.1$ &  &  0.01\ \ \ \ 0.00\ \ \ \ 0.00 & 1.00\ \ \ \ 1.00\ \ \ \ 1.00 \\ \hline
    \end{tabular}
    \caption{A summary of the effects of the cleaning procedure described in Section~\ref{sec:cleaning} for three choices of $\kappa=1.2, 2.0, 20.0$. $f_{{\rm merged}}$ denotes the fraction of haloes within a specific mass bin (as listed in the first column) that are deleted as separate entities in the halo catalogue and merged onto a larger object. $f_{{\rm inc}}$ denotes the fraction of haloes in each mass bin whose mass is increased (relative to the default halo catalogue) after they are merged with the cleaned haloes. The numbers listed here are taken from the \cph{} \highbase{} simulation at $z=0.5$.}
    \label{tab:clean_table}
\end{table}

The green curve shows the modification to the \compaso{} halo mass function after applying the cleaning strategy described in Section~\ref{sec:cleaning} assuming $\kappa=20.0$. This is the most conservative of three parameter variations we examine in this paper, flagging only those haloes for cleaning that were once part of haloes with mass $\geq20\times$ their mass at $z=0.5$. The net result is a very small change with respect to the default \compaso{} case, only marginally shifting the mass function to higher masses on scales greater than $\sim10^{14} {\rm M}_\odot/h$. The $\kappa=2.0$ version (in blue) shows a much more dramatic change, particularly in the regime of low-mass groups and clusters, reducing the discrepancy relative to \rockstar{} to the 25\% level. Finally, we see no noticeable change in transitioning from $\kappa=2.0$ to $\kappa=1.2$ (in brown), where the two curves overlap almost exactly. 

In Table~\ref{tab:clean_table}, we contrast the fraction of haloes (as a function of mass) that are affected by the cleaning method for each choice of $\kappa$, either because they are removed from the \compaso{} catalogue and merged, or because their reported mass has been augmented by the addition of merged neighbouring haloes. We see that, as expected, low-mass haloes are the ones that are more likely to be flagged for removal, while a larger fraction of the more massive haloes end up with increased masses after being merged with several smaller haloes. Furthermore, we see that the $\kappa=1.2$ and $2.0$ cases yield very similar results, consistent with what we see in Figure~\ref{fig:hmf_compare}; on the other hand, the choice of $\kappa=20.0$ tends to be far less effective at removing intermediate mass haloes.

\begin{figure*}
    \centering
    \includegraphics[width=\textwidth,trim={1in 1in 0 0}]{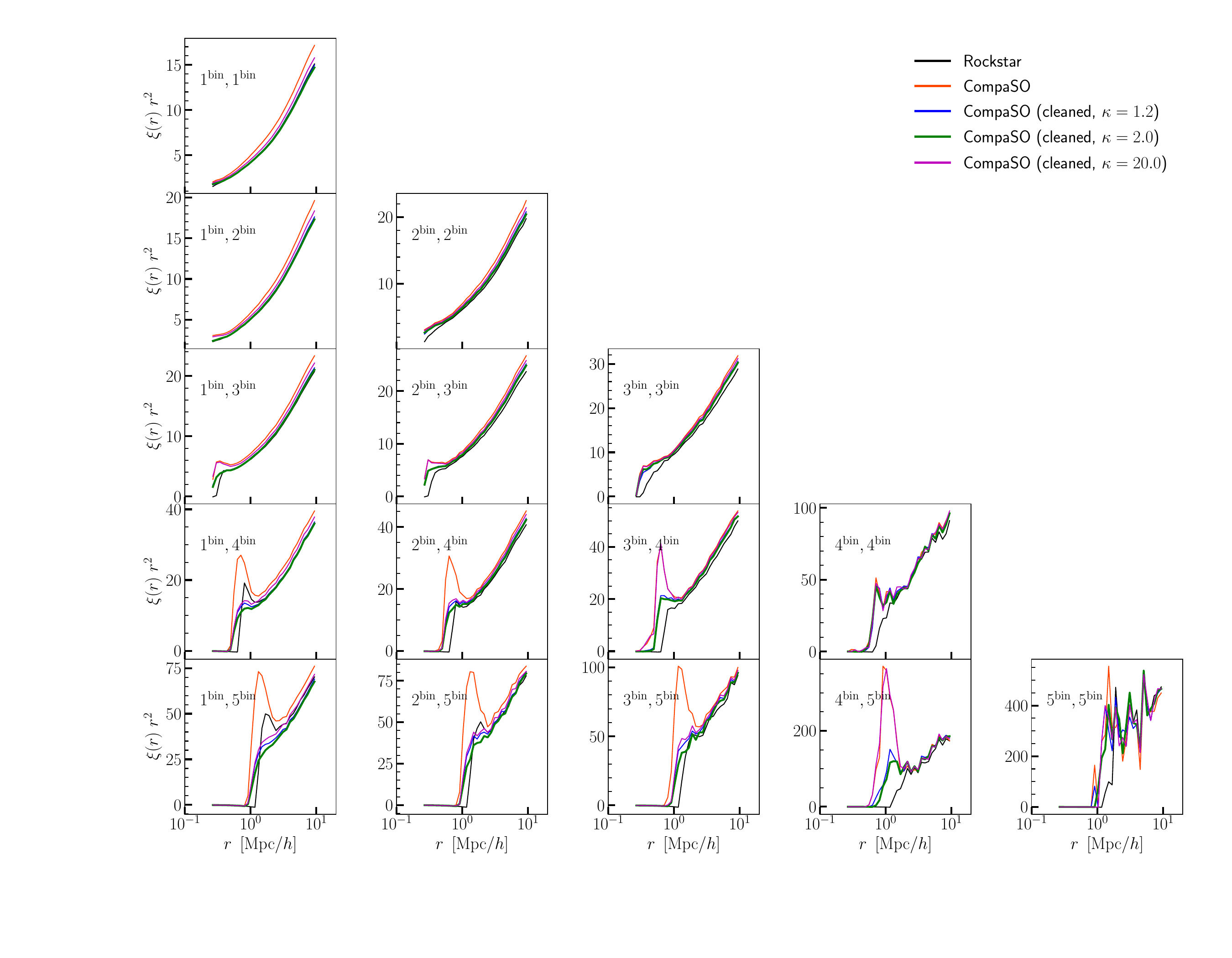}
    \caption{The two-point clustering of haloes at $z=0.5$ measured using \rockstar{} (black), 
\compaso{} (red) and the `cleaned' \compaso{} (with $\kappa=1.2$, blue; $\kappa=2.0$, green; $\kappa=20.0$, magenta) catalogues for the \highbase{} \cph{} simulation. Haloes are split into 5 distinct mass bins: $\log\left[M_{{\rm halo}}/ {\rm M}_\odot/h \right] = 11.5\pm 0.1$ (first bin); $\log\left[M_{{\rm halo}}/ {\rm M}_\odot/h \right] = 12.0\pm 0.1$ (second bin); $\log\left[M_{{\rm halo}}/ {\rm M}_\odot/h \right] = 12.5\pm 0.1$ (third bin); $\log\left[M_{{\rm halo}}/{\rm M}_\odot/h \right] = 13.5\pm 0.1$ (fourth bin); $\log\left[M_{{\rm halo}}/ {\rm M}_\odot/h \right] = 14.5\pm 0.3$ (fifth bin). Individual panels show the auto- and cross-correlation functions of haloes identified between pairs of mass bins. The unprocessed \compaso{} catalogue shows a strong excess of low-mass haloes in the outskirts of galaxy clusters, which manifests as a sharp peak in the cross-correlation function on scales $\sim 1$Mpc$/h$, well in excess of what is measured for \rockstar{} haloes. Cleaning the halo catalogue using the strategy described in Section~\ref{sec:cleaning} brings \compaso{} into much closer agreement with \rockstar{}, although there are residual differences in the highest mass bins. We also find that, as in the case of the halo mass function (Figure~\ref{fig:hmf_compare}), the difference between the $\kappa=1.2$ and $\kappa=2.0$ catalogues is almost negligible. The $\kappa=20.0$ case, which results in the least aggressive cleaning, shows marked differences and yields results that are more comparable to the default \compaso{} haloes.}
    \label{fig:corr_kappa2p0}
\end{figure*}

% \begin{figure*}
%     \centering
%     \includegraphics[width=\textwidth,trim={0 1in 0 0}]{Figures/corr_many_compare_allkappa.pdf}
%     \caption{As Figure~\ref{fig:corr_kappa2p0}, but now showing the {\it fractional difference} (relative to `uncleaned' \compaso{}) in the two-point correlation functions of haloes for the `cleaned' catalogues obtained assuming $\kappa=1.2$ (red), $\kappa=2.0$ (blue), and $\kappa=20.0$ (green). The ordering of the mass bins is identical to that in Figure~\ref{fig:corr_kappa2p0}. We find that, as in the case of the halo mass function (Figure~\ref{fig:hmf_compare}), the difference between the $\kappa=1.2$ and $\kappa=2.0$ catalogues is almost negligible. The $\kappa=20.0$ case, which results in the least aggressive cleaning, shows marked differences and yields results that are more comparable to the default \compaso{} haloes.}
%     \label{fig:corr_kappa_compare}
% \end{figure*}

\begin{figure}
    \centering
    \includegraphics[width=\columnwidth]{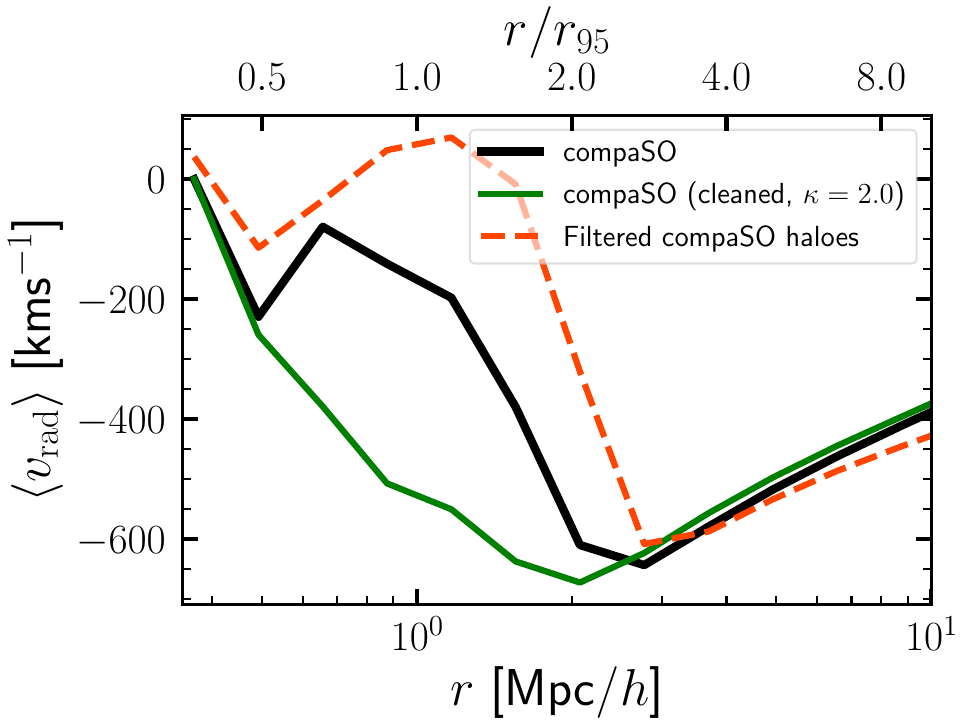}
    \caption{The mean peculiar radial velocity, $\left<V_{{\rm rad}}\right>$, of haloes in the mass range $\log\left[ M_{{\rm halo}}/{\rm M}_\odot/h\right]=\left[11.0,13.0\right]$, relative to a sample of low-mass cluster haloes in the mass range $\log\left[ M_{{\rm halo}}/{\rm M}_\odot/h\right]=\left[14.0,14.5\right]$. The curves in black and green, respectively, show results from the \compaso{} and cleaned \compaso{} catalogues, where we assume $\kappa=2.0$. The dashed red curve shows the radial profile for only those haloes that have been flagged and removed due to cleaning (i.e., merged into a larger halo) according to the procedure described in Section~\ref{sec:cleaning}. The upper $x$-axis denotes the radial range in units of $r_{95}$ from the cluster centre, where $r_{95}$ is the radius encompassing 95\% of the mass of the halo. The cleaning tends to remove haloes with more positive radial velocities in and around the outskirts of clusters (1-3 Mpc$/h$), with a sizeable population with $\left<v_{{\rm rad}}\right> >0$. This is likely representative of population of haloes that have exited their host clusters following at least one orbital passage.}
    \label{fig:radial_vel}
\end{figure}

\subsection{Length of the main progenitor branch}
\label{sec:mainbranch}
In the merger tree  comparison project presented in \cite{Srisawat2013}, a particularly useful metric for comparing different algorithms turns out to be a measure of the length of the main progenitor branch of haloes -- essentially, the number of snapshots into the past that one is able to reliably track a halo's main progenitor.

Figure~\ref{fig:length_branch_compare} shows the results of measuring the length of the main progenitor branch for the default (black) and cleaned (green) \compaso{} catalogues. To perform the cleaning, we have assumed our fiducial value of $\kappa=2.0$. The two curves largely overlap, suggesting that cleaning does not change this measure significantly, which is to be expected since the objects that are affected by cleaning comprise a small fraction of the full population of haloes (see Table~\ref{tab:clean_table}). The most prominent difference is the sharp reduction haloes that can only be tracked for 0 or 1 snapshots in the cleaned catalogue. This is dominated by objects in the default \compaso{} catalogue that have been flagged and identified for merging by the algorithm in Section~\ref{sec:cleaning}. This fraction is larger in the lower mass bins ($N_p\geq500$), as expected.

From Figure~\ref{fig:length_branch_compare} we also note that haloes that are better resolved are more easily tracked through multiple output times, with the vast majority of haloes ($\gtrsim 70\%$) with $N_p\geq1000$ being tracked across the entire length of the merger tree (denoted by the dashed vertical line). Finally, we find that in the lowest mass bin considered ($50\leq N_p\leq100$) it becomes increasingly difficult to track haloes through several snapshots. The reason for this is that reliably associating haloes across snapshots becomes unstable when only a few particles are available for matching. It is also limited by the choice we make in the construction of the merger tree in which we require that a halo contain at least 5 subsampled particles in order to be considered for tracking. An increasing fraction of haloes fail this test as we proceed to earlier snapshots.

An interesting way to further assess the fidelity of the merger trees for cleaned versus uncleaned haloes would be through visual characterisation, such as the dendogram diagnostic proposed by \cite{Poulton2018}. We leave such an exercise for future study.

\subsection{2-point clustering: the auto- and cross-correlation functions of haloes}
\label{sec:clustering}

In the previous subsections, we have seen that applying our \compaso{} catalogue cleaning procedure has the effect of flagging low-mass objects ($\log\left[ M_{{\rm halo}} / {\rm M}_\odot/h \lesssim14.5\right]$) onto larger, neighbouring haloes. This suggests that a large number of ``offending'' haloes (i.e., those that are flagged and merged by our cleaning algorithm) corresponds to low-mass haloes that exist in the peripheries of more massive haloes. This assertion can be tested by examining the clustering of haloes, measured using the 2-point correlation function. 

We calculate the correlation function, $\xi(r)$, using the \cite{Peebles1980} estimator:
\begin{equation*}
 \xi(r) = \frac{\left< DD(r) \right>}{\left< RR(r) \right>} - 1,
\end{equation*}
where $\left< DD(r)\right>$ and $\left< RR(r)\right>$, respectively, are the normalised data-data and random-random pair counts of objects separated by a distance $r$. To examine the clustering as a function of mass, we split the halo population into 5 mass bins: $\log\left[M_{{\rm halo}}/ {\rm M}_\odot/h \right] = 11.5\pm 0.1$; $12.0\pm 0.1$; $12.5\pm 0.1$; $13.5\pm 0.1$; $14.5\pm 0.3$. In practice, we compute $\xi(r)$ using the {\sc Corrfunc} package \citep{Sinha2020}.

Figure~\ref{fig:corr_kappa2p0} shows the bin-by-bin auto-correlation (diagonal panels) and cross-correlation (off-diagonal panels) functions of haloes identified by \rockstar{} (black), \compaso{} (red), as well as the cleaned \compaso{} catalogue obtained using our recommended choice of $\kappa=2.0$ (green), as well as $\kappa=1.2$ (blue), and $\kappa=20.0$ (magenta). The clustering is measured at $z=0.5$ in the \highbase{} \cph{} simulation. In each case, we have taken care to ensure that the clustering is computed at {\it the same halo abundance}; this guarantees that any systematic differences measured in $\xi(r)$ are as a result of {\it physical} differences in the three halo catalogues, rather than being due to a {\it different number} of objects being included based on the mass definition used in each case. 

Upon comparing the clustering in the default \compaso{} catalogue with \rockstar{}, we immediately notice significant differences. It is clear that \compaso{} has a greater tendency to find low-mass dark matter haloes and identify them as independent objects in their own right. This manifests first as a larger amplitude in the auto-correlation function of $\log\left[M_{{\rm halo}}/ {\rm M}_\odot/h \right] = 11.5$ and more prominently as the sharp spike in the cross-correlations of low-mass haloes with heavier objects. In particular, the spike appears around the transition region between the one- and two-halo terms in the cross-correlation functions ($r\approx1-3$ Mpc$/h$), which corresponds roughly to the typical virial radii of the most massive clusters in our sample. \rockstar{}, on the other hand, does not show such a pronounced feature in the same regime. In other words, the \compaso{} method finds an excess of low-mass dark matter structures floating around the boundaries of massive haloes; \rockstar{} seemingly does not deblend these sets into populations of independent objects.

Figure~\ref{fig:corr_kappa2p0} also shows that much of the discrepancy between \rockstar{} and \compaso{} disappears after the cleaning procedure is applied to the \compaso{} haloes. By merging haloes that our method flagged for cleaning with heavier neighbours, we eradicate the vast majority of the population of haloes that were originally found at the boundaries of massive clusters. This results in a dramatic suppression in the prominent spike that previously punctuated the measurement of $\xi(r)$ for \compaso{} haloes. That said, the auto-correlation functions of haloes heavier than $\log\left[M_{{\rm halo}}/ {\rm M}_\odot/h \right] \geq 12.5$ remain largely unaffected by cleaning, and they show appreciable differences with respect to \rockstar{} on scales smaller than $r\lesssim2$ Mpc$/h$. We note, however, that matching \rockstar{} exactly is not our intention -- nor, indeed, is it necessary -- as some differences between the two catalogues will inevitably persist due to the different philosophies used for identifying dark matter haloes.

%In Figure~\ref{fig:corr_kappa_compare} we show the dependence of the 2-point correlation function on the particular choice of $\kappa$. Each panel now shows the {\it fractional change} in $\xi(r)$ (measured relative to the default \compaso{} measurement) for $\kappa=1.2, 2.0$, and $20.0$. 
We also find that, consistent with our observations of the halo mass function in Figure~\ref{fig:hmf_compare}, $\kappa=1.2$ and $\kappa=2.0$ yield identical results in all panels. Both choices lead to a $\sim15\%$ reduction in the auto-correlation of $\log\left[M_{{\rm halo}}/ {\rm M}_\odot/h \right] \leq 12.0$ haloes. The $\kappa=20.0$ case results in a more modest change ($5-10\%$ in the auto-correlation function). The biggest contrast to the lower $\kappa$ (i.e., more aggressive) cleaning choices is seen in the cross-correlations between massive haloes. This is seen most prominently in bins (3,4) and (4,5) of Figure~\ref{fig:corr_kappa2p0}, where we see that the $\kappa=20.0$ choice has no real cleaning effect relative to the default \compaso{} halo catalogue. 

\subsection{Radial infall velocity profiles}
\label{sec:infall}

Thus far, we have seen that the population of haloes that is preferentially removed by our cleaning method corresponds to those that are typically located at the boundaries of massive clusters. Their presence manifests as a sharp bump in the cross-correlation function on the scale of $1-3$ Mpc$/h$, which is comparable to the typical virial radius of low-mass clusters. Next, we examine the dynamics of the `cleaned' objects as measured by their infall velocities in the vicinity of the neighbouring clusters.

Figure~\ref{fig:radial_vel} shows the mean (peculiar) radial velocity profile of a tracer population of haloes, measured relative to clusters in the mass range $\log\left[ M_{{\rm halo}}/{\rm M}_\odot/h\right]=\left[14.0,14.5\right]$. For the tracer population, we select haloes in the mass range $\log\left[ M_{{\rm halo}}/{\rm M}_\odot/h\right]=\left[11.0,13.0\right]$. The profiles are shown for the default \compaso{} (black) and cleaned \compaso{} (with $\kappa=2.0$, green) catalogues, both measured at $z=0.5$ in the \highbase{} \cph{} simulation. The peculiar radial velocity, $\left< v_{{\rm rad}}\right>$, is measured as a function of distance from the cluster centre. The upper axis denotes this distance in units of $r_{95}$, which is the radius that encompasses 95\% of the mass of the cluster measured from its L2 centre-of-mass. 

First, we see that the radial velocity profiles measured in both the default and cleaned \compaso{} cases show similar overall trends. As haloes get increasingly close to the cluster centre from $r\sim10$ Mpc$/h$ to about $r\sim3$ Mpc$/h$, $\left<v_{{\rm rad}}\right>$ becomes increasingly negative, signifying the fact that these objects are infalling into the cluster. As these objects reach the vicinity of the virial radius of the cluster (comparable to the value of $r_{95}$), there is a sharp upturn in the profile, signifying that the velocities of the tracer haloes begin to virialise within the cluster. The system then reaches equilibrium in the innermost portion of the halo, where $\left<v_{{\rm rad}} \right>=0$.

The behaviour of the \compaso{} haloes before and after cleaning are identical on scales larger than about 3 Mpc$/h$. Within this radius, however, there is a systematic deviation in the profile, with the haloes in the cleaned catalogue showing a more negative infall velocity on average than in the default \compaso{} case. Notably, this change is seen in and around the virial radius of the cluster haloes, which is indeed the scale where the cross-correlation function of haloes in each catalogue differed the most. 

We can understand the origin of this shift by considering the mean radial infall velocity profile of those objects that have been flagged by our cleaning procedure and removed from the \compaso{} catalogue; in Figure~\ref{fig:radial_vel}, this is shown as the dashed red curve. The profile for the filtered objects is shifted systematically to more positive values of $\left<v_{{\rm rad}}\right>$. It is especially noticeable that $\left<v_{{\rm rad}}\right>>0$ in the regime between 0.8-2$\, r_{95}$, which is indicative of a population of haloes whose radial motions point away from the cluster centre. These are likely examples of ``splashback'' haloes -- those that were once part of a larger object, but have since exited following at least one orbital passage within their former hosts. The cleaning method we have applied guarantees that the past merger tree of the filtered objects once shared a main progenitor with their neighbouring massive halo, which adds further confidence to our notion that the red curve in Figure~\ref{fig:radial_vel} receives a large contribution from these splashback objects. By removing the contribution of these objects from the mean radial infall velocity profiles, we more easily isolate the population of objects in the boundaries of massive clusters that are actually infalling at present. This also has a small effect in changing the location of the upturn in the velocity profile, shifting from $\approx3$ Mpc$/h$ to around 2 Mpc$/h$ between the default and cleaned \compaso{} catalogues.

\subsection{From cleaned haloes to mock galaxy catalogues}
\label{sec:HOD}

As a final test of the value of our cleaning method, we apply a set of `cleaned' \summit{} simulations in a real use-case scenario: building mock galaxy catalogues to match the number density and redshift-space clustering of an observed galaxy sample. 

Specifically, we utilise galaxies from the CMASS sample of the Baryon  Oscillation Spectroscopic Survey \citep[BOSS;][]{Bolton2012,Dawson2013} Data Release 12, part of the Sloan Digital Sky Survey (SDSS) III programme \citep{Eisenstein2011}. In particular, we focus on galaxies in the redshift range $0.46<z<0.6$, yielding a sample of approximately 600,000 galaxies (predominantly Luminous Red Galaxies) at an average number density of $n_{{\rm gal}}=\left(3.01\pm 0.3 \right)\times 10^{-4}h^3$Mpc$^{-3}$. The redshift-space clustering of the sample is determined by the anisotropic 2-point correlation function, $\xi\left(r_p,\upi\right)$. $r_p$ and $\upi$, respectively, are the transverse and line-of-sight galaxy separation in comoving units. We utilise the full-shape anisotropic clustering instead of the projected clustering as it provides stronger constraints on the HOD \citep{Yuan2021}. 
We determine the anisotropic 2-point correlation function $\xi\left(r_p,\upi\right)$ using the \cite{Landy1993} estimator:
\begin{equation*}
    \xi\left(r_p,\upi\right) = \frac{DD-2DR+RR}{RR}\;,
\end{equation*}
where $DD$, $DR$, and $RR$ are the data-data, data-random, and random-random pair counts in bins of $\left(r_p,\upi \right)$. For the present analysis, we choose 8 logarithmically-spaced bins between 0.169 Mpc$/h$ and 30 Mpc$/h$ in the transverse direction, and 6 linearly-spaced bins between 0 and 30 Mpc$/h$ bins along the line-of-sight direction. 

To produce the galaxy mock, we make use of the \summit{} \textsf{Base\_c000\_ph000} simulation, which provides 8$\times$ larger volume than the \highbase{} box we have considered until now. The mass resolution ($\sim2\times10^9\,{\rm M}_\odot/h$ per particle) is more than sufficient for the CMASS galaxies, which typically live in haloes of $> 10^{12}\,{\rm M}_\odot/h$.
We populate \compaso{} dark matter haloes with galaxies using the \textsc{AbacusHOD}\footnote{\url{https://abacusutils.readthedocs.io/en/latest/hod.html}} code \citep{Yuan2022}, which determines the galaxy-halo connection using sophisticated particle-based HOD models. Specifically, the HOD we use for this analysis is based on the \cite{Zheng2007} model where, respectively, the mean central and satellite occupations in haloes with mass $M_h$ are defined as:
\begin{equation}\label{eq:ncen}
    \left< N_{{\rm cen}} (M_h) \right> = \frac{1}{2} {\rm erf} \left[ \frac{\log\left(M_{{\rm cut}} / M_h\right) }{\sqrt{2}\sigma} \right] \;,
\end{equation}
and
\begin{figure*}
    \centering
    \includegraphics[width=\textwidth]{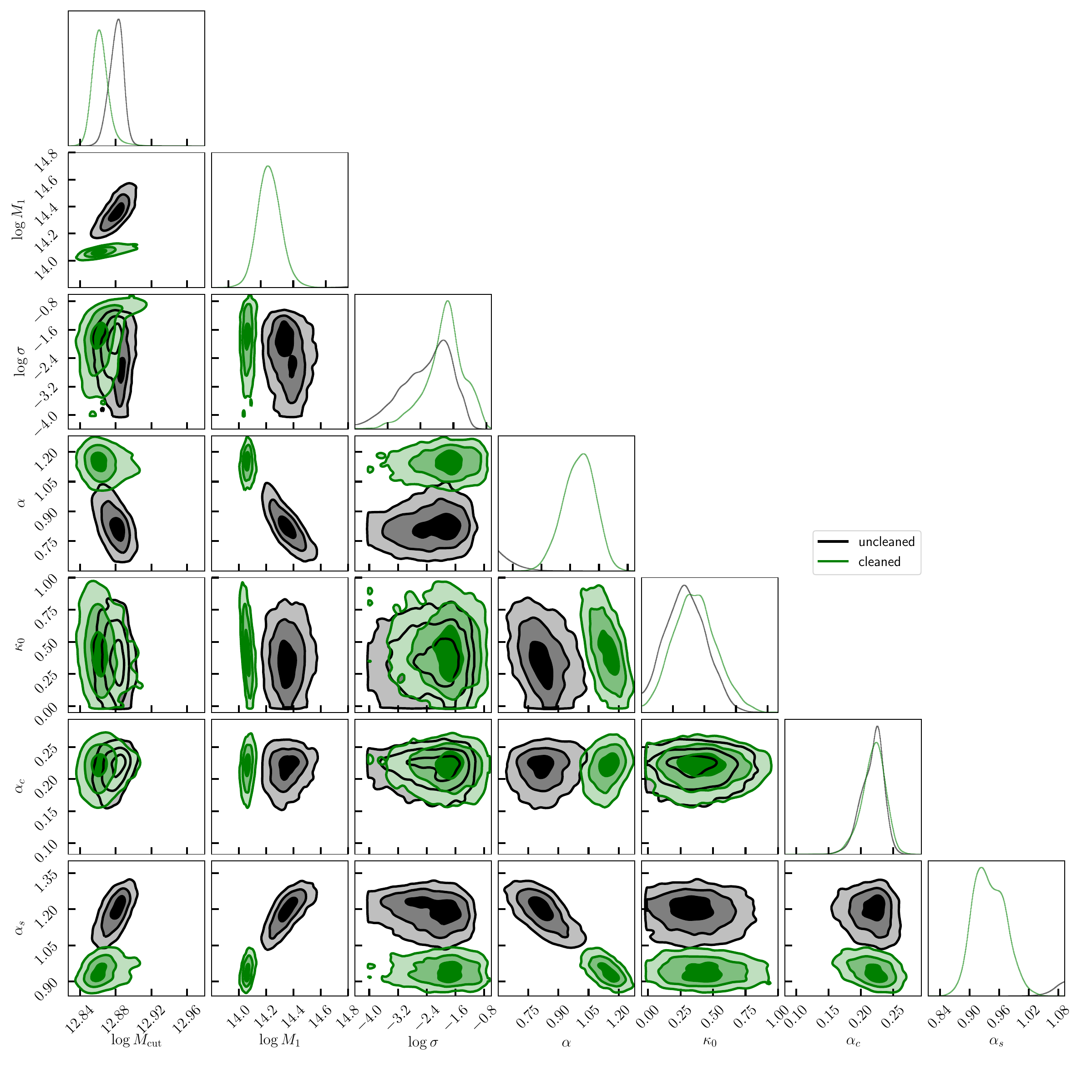}
    \vspace{-15pt}
    \caption{The 1D and 2D marginalised posterior constraints on our 7-parameter HOD fits to the BOSS CMASS DR12 redshift-space 2-point correlation function, $\xi\left(r_p,\upi\right)$. The distributions for the uncleaned and cleaned halo catalogues, respectively, are represented by the black and green contours. The three levels correspond to $1,2,$ and $3\sigma$ level constraints. The posterior medians on each of these parameters are listed in the first 7 rows of Table~\ref{tab:HOD_fits}.}
    \label{fig:posteriors}
\end{figure*}
\begin{equation}\label{eq:nsat}
    \left<N_{{\rm sat}} (M_h)\right> = \left( \frac{M_h-\kappa_0 M_{{\rm cut}}}{M_1} \right)^\alpha \left< N_{{\rm cen}} (M_h) \right>\;.
\end{equation}
Here, $M_{{\rm cut}}$, $\sigma$, $\kappa_0$, $M_1$, and $\alpha$ are model parameters determined through fitting. $M_{{\rm cut}}$ is the minimum mass of haloes that host a central galaxy, and $\sigma$ is the width of the transition from non-zero to zero central occupancy. Furthermore, $M_1$ is the characteristic mass of haloes that host satellites, while $\kappa_0 M_{{\rm cut}}$ represents the minimum halo mass for hosting at least one satellite. $\alpha$ is the power-law slope of the satellite occupation. Galaxy positions in \textsc{AbacusHOD} are assigned such that the central galaxy is located at the host halo centre, while satellites are placed on halo particles. Note that a more common notation for $\kappa_0$ is simply $\kappa$; we have chosen the former to avoid confusion with the cleaning factor, $\kappa$, we have adopted until now. 

Because of redshift-space distortions, the line-of-sight velocities of the galaxies are translated into biases in the measurements of galaxy positions. It is thus important to have a model of the galaxy velocities, particularly their line-of-sight component. 
In this analysis, we allow for velocity bias effects, which introduces flexibilities in the velocities of the central and satellite galaxies \citep{2015aGuo}. Specifically, the central galaxy velocity along the line-of-sight is given by:
\begin{equation}
    v_\mathrm{cent, z} = v_\mathrm{L2, z} + \alpha_c \delta v(\sigma_{\mathrm{LOS}}),
    \label{equ:alphac}
\end{equation}
where $v_\mathrm{L2, z}$ denotes the line-of-sight component of the L2 centre-of-mass velocity, $\delta v(\sigma_{\mathrm{LOS}})$ denotes the Gaussian scatter, and $\alpha_c$ is the central velocity bias parameter, which modulates the amplitude of the central velocity bias effect. By definition, $\alpha_c$ is non-negative, and $\alpha_c = 0$ corresponds to no central velocity bias, in which case the central galaxy simply assumes the centre-of-mass velocity of the L2 subhalo.
The satellite velocity is given by 
\begin{equation}
    v_\mathrm{sat, z} = v_\mathrm{L2, z} + \alpha_s (v_\mathrm{p, z} - v_\mathrm{L2, z}),
    \label{equ:alpha_s}
\end{equation}
where $v_\mathrm{p, z}$ denotes the line-of-sight component of particle velocity, and $\alpha_s$ is the satellite velocity bias parameter. $\alpha_s = 1$ corresponds to no satellite velocity bias, in which case the satellite simply assumes the velocity of the particle.
\begin{figure*}
    \centering
    \includegraphics[width=0.48\textwidth]{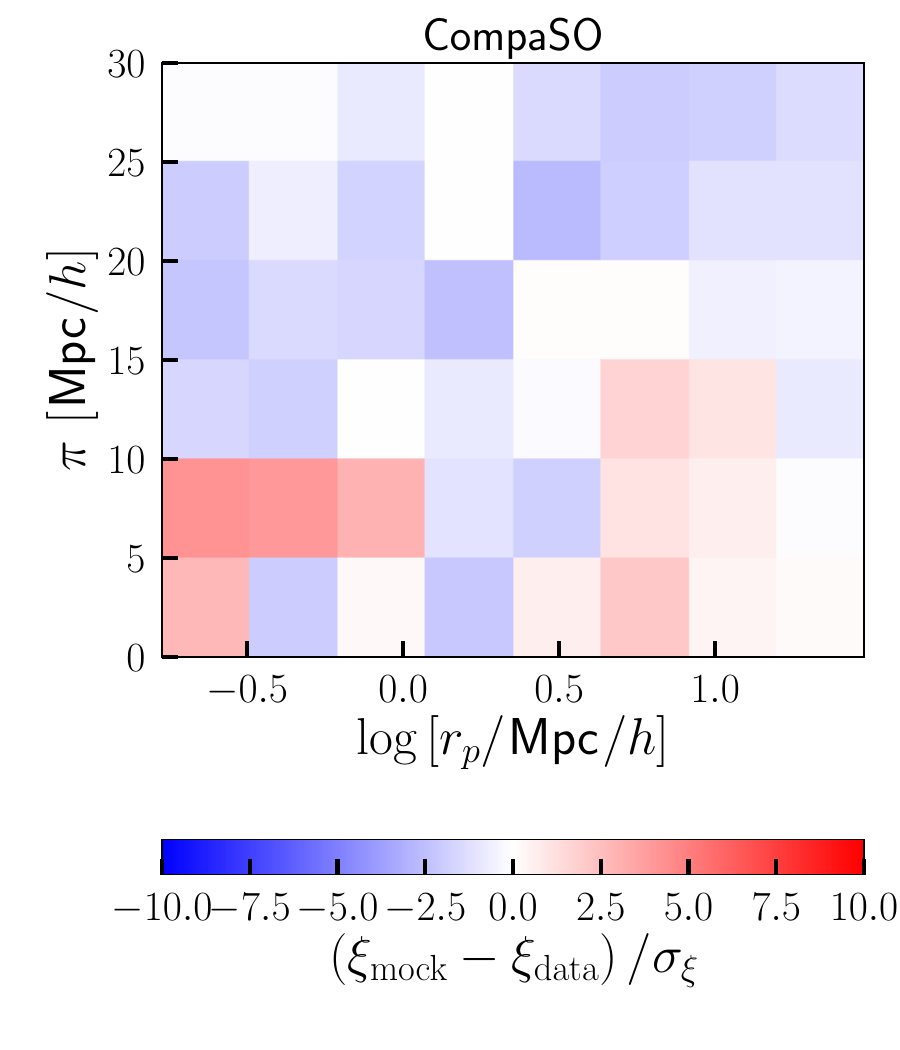}
    \includegraphics[width=0.48\textwidth]{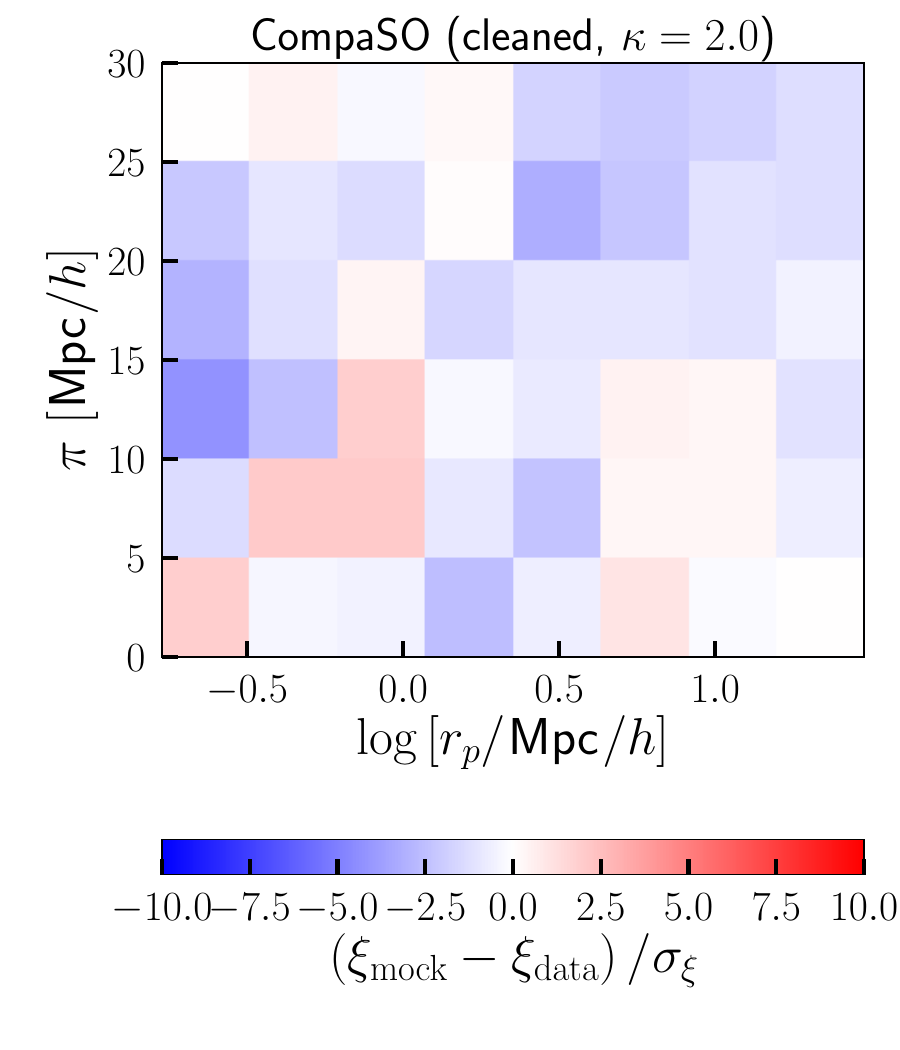}
    \caption{The best-fit 2-point correlation function fits to $\xi\left(r_p,\upi\right)$ using the uncleaned (left) and cleaned (right) versions of the \textsf{Base\_c000\_ph000} simulation. The colour maps show the {\it residual} on the best-fitting mocks with respect to the data. The errors, $\sigma_\xi$ are computed using the diagonal elements of the inverse covariance matrix (see main text). The mocks constructed using the cleaned \compaso{} catalogues yield a better fit to the clustering, particularly in the lower left and lower right quadrants.}
    \label{fig:xi_bestfit}
\end{figure*}
In summary, our HOD model consists of 7 parameters: 5 in the baseline HOD model ($M_\mathrm{cut}, M_1, \sigma, \alpha, \kappa_0$), and 2 velocity bias parameters ($\alpha_c, \alpha_s$).
\textsc{AbacusHOD} is also capable of implementing multi-tracer occupation, and presents additional important extensions to the baseline HOD model including assembly bias and satellite profile flexibility \citep[see][for details]{Yuan2018}. We do not incorporate these extensions in the present work. A full analysis enabling all relevant extensions is described in \cite{Yuan2022}.

In order to determine the best-fit values of the free parameters in Eqs.~(\ref{eq:ncen}) and~(\ref{eq:nsat}), we perform a joint fit of the CMASS DR12 galaxy number density, $n_{{\rm gal}}$, and the redshift-space 2-point correlation function, $\xi\left(r_p,\upi\right)$. The quality of the fit is determined using a chi-square statistic, defined as:
\begin{equation*}
    \chi^2 = \chi_\xi^2 + \chi_{n_{{\rm gal}}}^2
\end{equation*}
where
\begin{equation*}
       \chi^2_{\xi}  = (\bm{\xi}_{\rm{mock}} - \bm{\xi}_{\rm{data}})^T \bm{C}^{-1}(\bm{\xi}_{\rm{mock}} - \bm{\xi}_{\mathrm{data}}),
\end{equation*}
and 
\begin{equation*}
   \chi^2_{n_g} = \begin{cases}
   \left(\frac{n_{\mathrm{mock}} - n_{\mathrm{data}}}{\sigma_{n}}\right)^2 & (n_{\mathrm{mock}} < n_{\mathrm{data}}) \\
   0 & (n_{\mathrm{mock}} \geq n_{\mathrm{data}}).
   \end{cases}
   \label{equ:chi2ng}
\end{equation*}

Here, $\bm{C}$ is the jackknife covariance on $\xi$, and $\sigma_n$ is the jackknife uncertainty on the galaxy number density. $f_{{\rm ic}}$ is an incompleteness factor that uniformly downsamples our mocks to match the desired number density. We sample the posterior space with the \textsc{dynesty} nested sampling code \citep{Speagle2020}. Full details of this fitting procedure are described in Section 4.1 of \cite{Yuan2021}.

Table~\ref{tab:HOD_fits} demonstrates the significant improvement in the fit to $\xi\left(r_p,\upi\right)$ due to cleaning. Specifically, the two columns summarise the best-fit HODs obtained using mocks constructed from the default (uncleaned) \compaso{} halo catalogues (left column) and from the `cleaned' version of this catalogue assuming $\kappa=2.0$ (right column). The bottom two rows compare the the log-likelihood, $\log \mathcal{L}$, and Bayesian evidence, $\log \mathcal{Z}$, resulting from each of the cleaned and uncleaned catalogues. The Bayesian evidence, which describes the consistency between the model and the data, for the uncleaned mock catalogue is $\log\mathcal{Z}=-76$, compared to $\log\mathcal{Z}=-62$ for the cleaned version. In other words, the BOSS redshift-space clustering exhibits a roughly 14 $e$-fold preference for the cleaned mocks over the uncleaned ones. Similarly, we see a 14 $e$-fold improvement to the best-fit likelihood. 

The full marginalised posterior distributions of our 7-parameter HOD mocks fit to $\xi\left(r_p,\upi\right)$ are shown in Figure~\ref{fig:posteriors}, in which the parameter constraints for the uncleaned catalogue are shown in black, while for the cleaned catalogue these are in green. The maximum likelihood values of the HOD parameters are listed in the first 7 rows of Table~\ref{tab:HOD_fits}. Unsurprisingly, the switch from uncleaned to cleaned \compaso{} haloes results in a shift in the best-fit HOD parameters, particularly $M_1$, $\alpha$, and $\alpha_s$. The decrease in $M_1$ and the increase in $\alpha$ suggest a larger satellite population in the cleaned mock, consistent with the fact that the halo cleaning re-merges smaller haloes on the fringe of larger haloes, converting centrals in these fringe haloes to satellites in the re-merged halo. The decrease in the satellite velocity bias parameter, $\alpha_s$, pulls it back into agreement with no satellite velocity bias ($\alpha_s = 1$). This is consistent with prior velocity bias fits by e.g. \cite{2015aGuo}. Judging from the marginalised posterior contours, it is also worth noting that the cleaning also generally yields better-constrained HOD parameter posteriors and the removal of several parameter degeneracies. This suggests that the cleaned haloes are indeed more consistent with the data, and thus results in a preferred forward model.
\begin{table}
    \centering
    \begin{tabular}{c|c|c}
    \hline \hline
         & \compaso{} & \compaso{} (cleaned, $\kappa=2.0$) \\
    \hline 
    $M_{{\rm cut}}$ & 12.887 & 12.857 \\
    $M_1$ & 14.387 & 14.055 \\
    $\log \sigma$ & -2.651 & -1.807 \\
    $\alpha$ & 0.813 & 1.189 \\
    $\kappa_0$ & 0.335 & 0.343 \\
    $\alpha_c$ & 0.226 & 0.227 \\
    $\alpha_s$ & 1.217 & 0.913 \\
    \hline
     $\log \mathcal{Z}$ & -76  & {\bf -62} \\
     $\log \mathcal{L}$ & -57 & {\bf -43} \\
     \hline
    \end{tabular}
    \caption{A summary of the best-fit HOD parameters (first 7 rows) derived from mocks built on the uncleaned (left column) and cleaned (righ column) versions of the \textsf{Base\_c000\_ph000} simulation, assuming $\kappa=2.0$. The last two rows, respectively, list the Bayesian evidence, $\log\mathcal{Z}$,  and log-likelihood, $\log\mathcal{L}$, of the best-fit mocks. The corresponding values suggest a significant preference for the cleaned catalogue over the uncleaned version.}
    \label{tab:HOD_fits}
\end{table}
Finally, Figure~\ref{fig:xi_bestfit} shows the residual on the best-fit redshift-space 2-point correlation function, $\xi\left(r_p,\upi\right)$ obtained using mocks constructed from the uncleaned halo catalogues (left panel) and from the cleaned catalogue (right panel). In particular, there is a noticeable reduction in the residual on scales $r_p\sim3-10$ Mpc$/h$, where the uncleaned \compaso{} mocks predict an excess in clustering relative to the data. The removal of low-mass, high velocity haloes at the boundaries of groups and clusters (see Figure~\ref{fig:radial_vel}) in the cleaned version results in an improved fit. The shift from uncleaned to cleaned haloes also results in an appreciable improvement in the residuals on sub-Mpc projected separations. While it is difficult to conduct `$\chi$ by eye', the right panel sums up to significantly lower $\chi^2$ compared to the left panel (see $\log \mathcal{L}$ comparison in Table~\ref{tab:HOD_fits}). 

\section{Conclusions}
\label{sec:conclusions}

Tracking the formation and growth of dark matter haloes is a critical step in the analysis of cosmological simulations of structure formation. In this paper, we have introduced an algorithm for constructing high-fidelity halo merger trees from \summit{} \citep{Maksimova2021}, an extremely large suite of cosmological simulations that has been designed to meet the Cosmological Simulation Requirements of the Dark Energy Spectroscopic Instrument (DESI) survey. The flagship \summit{} dataset consists of 139 simulations of size 2 Gpc$/h$, each resolved with 6912$^3$ particles, yielding an effective particle mass of $2\times10^9 {\rm M}_\odot/h$, making it an ideal dataset for theoretical applications in the mass scale relevant to emission line galaxies (ELGs), a primary target for DESI. 
Combined with the several smaller and larger boxes that we have also simulated as part of this project, \summit{} totals nearly 60 trillion particles, with simulations spanning a wide range of cosmological parameter variations, with the majority also including the effect of massive neutrinos on structure growth.

To identify haloes in \summit{}, we use the {\it Competitive Assignment to Spherical Overdensities}, or \compaso{}, algorithm \citep{Hadzhiyska2021}, which is a highly-efficient, on-the-fly halo finder that has been purpose-built for \summit{}. \compaso{} identifies haloes from a given particle set by first applying a density-bounded friends-of-friends linking algorithm to identify groups, followed by a hierarchical sequence of particle assignment steps that competitively link together particles that satisfy certain criteria based on pre-defined density thresholds. A brief outline of how \compaso{} operates is provided in Section~\ref{sec:compaso}.

Like several other merger tree algorithms that have been published in the literature \citep[e.g.][]{Behroozi2013b,Jiang2014,Poole2017,Han2018,Elahi2019,Rangel2020}, in our method, which is described in detail in Section~\ref{sec:algorithm}, the ability to build associations between sets of haloes relies on the ability to accurately track the particle content of these entities across multiple snapshots. In performing these matches, particular preference is given to  particles with the highest kernel densities (as computed by \compaso{}). This tends to trace material in the cores of haloes, which has been shown to be more reliable choice for the construction of merger trees \citep[e.g.][]{Srisawat2013}. A potential halo association occurs when some fraction of these particles overlaps between any two haloes across two adjacent output times. Once these associations are established, we define halo progenitors as only those objects at some snapshot {\tt i-1} that contribute a sizeable fraction of matched particles (i.e., exceeding a threshold fraction) to a recipient halo at snapshot {\tt i}. A visualisation of our algorithm is shown in Figure~\ref{fig:algorithm}. In Section~\ref{sec:cadence}, we perform a series of tests to show that our algorithm is robust to the density of snapshots used to construct halo merger trees, particularly when measuring the kinds of statistics that are commonly used in the construction of mock catalogues (Figures~\ref{fig:cadence_mah}-\ref{fig:cadence_cmf}).

The most direct application of a merger tree is to compute the mass accretion histories of dark matter haloes. Given the mass resolution and volume of the \summit{} simulations presented in this paper, we are able to accurately track the assembly history of objects ranging from the ELG-mass hosts to rich clusters of galaxies (Figure~\ref{fig:mah_convergence}). We also find good convergence between pairs of simulations that have been run using the same initial phases, but that vary in resolution by a factor of 6. 

We then describe a procedure that uses merger trees to `clean' the default \compaso{} halo catalogues of objects whose masses are deemed to be unreliable due to physical and/or numerical effects during their past evolution. In particular, haloes identified at the present time may have undergone fly-bys, been `split' off larger haloes by the \compaso{} algorithm, or simply passed through the interiors of more massive objects. Each of these processes can hamper the accuracy with which the particle content (and, therefore, the present-day properties) of the halo are tracked, which can consequently bias models of the galaxy-halo connection that are applied to our default halo outputs. In \abacus{}, we take on a conservative approach and identify all such `problematic' objects and eliminate these as independent haloes using the procedure described in Section~\ref{sec:cleaning}. In short, this method flags those objects whose maximum mass (measured along the main progenitor branch) exceeds its present-day mass by some factor, $\kappa$. After identifying the more massive, neighbouring halo they were once part of, the objects are "merged" onto this host and the newly aggregated system is treated as a single entity at all subsequent output times (see Figures~\ref{fig:cleaningflow} \&~\ref{fig:clean_example}).

The net effect of this cleaning method is to predominantly flag low-mass haloes that have especially truncated merger trees (Figure~\ref{fig:length_branch_compare}), and those that are found preferentially near the boundaries of cluster-mass haloes. Their removal results in a suppression in the 2-point cross-correlation on scales of 1-3 Mpc$/h$ (Figure~\ref{fig:corr_kappa2p0}), and we find that the response of this effect is largely insensitive to small changes in $\kappa$. On examining the detailed dynamics of the `cleaned' haloes, we find that our method preferentially removes a large population of objects with net positive radial velocities directed away from the centres of clusters (Figure~\ref{fig:radial_vel}). 

As a final test, we apply our cleaned catalogues in a real use-case application of building Halo Occupation Distribution mocks to fit the BOSS CMASS DR12 redshift-space 2-point correlation function, $\xi\left(r_p,\upi\right)$. To this end, we use the \textsc{AbacusHOD} code \citep{Yuan2022} which allows for several generalised extensions to the baseline halo mass-only HOD model. We find that shifting from uncleaned to cleaned catalogues results in a enhanced satellite population in the latter, consistent with the idea low-mass haloes that was originally identified as individual entities are now merged with larger neighbours and become part of their satellite population (Figure~\ref{fig:posteriors}). We also find that the resulting fit to $\xi\left(r_p,\upi\right)$ is much improved in the cleaned version than in the uncleaned version, with Bayesian evidence and log-likelihood that suggest a 14 $e$-fold preference for the former over the latter (Table~\ref{tab:HOD_fits} and Figure~\ref{fig:xi_bestfit}). 

While in this paper we have presented results from only three sets of \summit{} simulations, our merger tree construction has been applied across the entire suite. This comprises an extensive repertoire of information pertaining to the redshift evolution of halo properties spanning a generous range in cosmology variations. This makes the \summit{} suite a potentially invaluable resource for constructing and testing models of the galaxy-halo connection, covariance estimation, and emulator building in anticipation of next generation galaxy redshift surveys.

\section*{Acknowledgements}

We thank the referee for providing a number of useful comments and suggestions that have strengthened quality of this work. We are grateful to Alex Smith for testing an early version of the halo catalogue cleaning method, and for identifying areas of improvement. This work has been supported by NSF AST-1313285, DOE-SC0013718, and NASA ROSES grant 12-EUCLID12-0004. SB is supported by the UK Research and Innovation (UKRI) Future Leaders Fellowship [grant number MR/V0233]. DJE is supported in part as a Simons Foundation investigator. NAM was supported in part as a NSF Graduate Research Fellow. LHG is supported by the Center for Computational Astrophysics at the Flatiron Institute, which is supported by the Simons Foundation. 

This research used resources of the Oak Ridge Leadership Computing Facility, which is a DOE Office of Science User Facility supported under Contract DE-AC05-00OR22725. Computation of the merger trees used resources of the National Energy Research Scientific Computing Center (NERSC), a U.S. Department of Energy Office of Science User Facility located at Lawrence Berkeley National Laboratory, operated under Contract No. DE-AC02-05CH11231. The \summit{} simulations have been supported by OLCF projects AST135 and AST145, the latter through the Department of Energy ALCC program.

We would like to thank the OLCF and NERSC staff for their highly responsive and expert assistance, both scientific and administrative, during the course of this project.
%%%%%%%%%%%%%%%%%%%%%%%%%%%%%%%%%%%%%%%%%%%%%%%%%%
\section*{Data Availability}

The \summit{} simulations have been placed into the public domain, subject to the academic citations described at \url{https://abacussummit.readthedocs.io/en/latest/citation.html}. A user interface for reading and analysing \summit{} data, as well as the \textsc{AbacusHOD} module, are available as part of the {\tt abacusutils} package (\url{https://abacusutils.readthedocs.io/en/latest/}).

Data access is available through OLCF's Constellation portal. The initial data release includes cleaned halo catalogues for 6 \summit{} simulations. The persistent DOI describing these data is \href{https://www.doi.org/10.13139/OLCF/1811689}{10.13139/OLCF/1811689}. Cleaned halo catalogues and particle subsamples for all other \summit{} boxes are available through the DOI \href{https://doi.ccs.ornl.gov/ui/doi/363}{10.13139/OLCF/1828535}. Instructions for accessing the data are given at \url{https://abacussummit.readthedocs.io/en/latest/data-access.html}.

%{\it Note to the editor and referee: In a subsequent DOI coordinated with OLCF's Constellation portal, we will release cleaned halo catalogues for all remaining \summit{} simulations, as well as the data for the \cadence{} simulation. These data sets are presently available to all users at NERSC, but not in a web-accessible location.} 

%%%%%%%%%%%%%%%%%%%% REFERENCES %%%%%%%%%%%%%%%%%%

% The best way to enter references is to use BibTeX:

\bibliographystyle{mnras}
\bibliography{references} % if your bibtex file is called example.bib

%%%%%%%%%%%%%%%%%%%%%%%%%%%%%%%%%%%%%%%%%%%%%%%%%%

% Don't change these lines
\bsp	% typesetting comment
\label{lastpage}
\end{document}